\documentclass{aa}  
\usepackage{graphicx}
\usepackage{txfonts}
\usepackage{xcolor}

\begin{document} 

   \title{Growing a nuclear star cluster from star formation and cluster mergers: The JWST NIRSpec view of NGC 4654\thanks{Based on observations with the NASA/ESA {\em Hubble Space Telescope} and the NASA/ESA/CSA {\em James Webb Space Telescope}, which are operated by AURA, Inc., under NASA contracts NAS5-26555 and NAS 5-03127.}}

   \subtitle{}
   \titlerunning{Growing a nuclear star cluster from star formation and young clusters}

   \author{Katja Fahrion\thanks{ESA research fellow}\fnmsep
          \inst{1}
          \and
          Torsten B\"oker
          \inst{2}
          \and
          Michele Perna
          \inst{3}
          \and
          Tracy L. Beck
          \inst{4}
          \and
        Roberto Maiolino
        \inst{5}\fnmsep\inst{6}
          \and
          Santiago Arribas
          \inst{3}          
          \and
          Andrew J. Bunker
          \inst{7}
          \and
          Stephane Charlot
          \inst{8}
          \and
          Matteo Ceci
          \inst{9}\fnmsep\inst{10}
          \and
          Giovanni Cresci
          \inst{10}
          \and
          Guido De Marchi
          \inst{1}
          \and
          Nora L\"utzgendorf
          \inst{1}
          \and 
          Lorenzo Ulivi
          \inst{9}\fnmsep\inst{10}\fnmsep\inst{11}
    }

   \institute{European Space Agency, European Space Research and Technology Centre, Keplerlaan 1, 2201 AZ Noordwijk, the Netherlands.\\
             \email{katja.fahrion@esa.int}
             \and %inst 2
             European Space Agency, c/o STScI, 3700 San Martin Drive, Baltimore, MD 21218, USA
             \and %inst 3
             %michele and Santiago
             Centro de Astrobiolog\'ia (CAB), CSIC--INTA, Departamento de Astrof\'\i sica, Cra. de Ajalvir Km.~4, 28850 -- Torrej\'on de Ardoz, Madrid, Spain
             \and %inst 4 Tracy
             Space Telescope Science Institute, 3700 San Martin Drive, Baltimore, MD 21218, USA 
            \and %inst 5 Roberto
             Cavendish Laboratory, University of Cambridge, 19 J. J. Thomson Ave., Cambridge CB3 0HE, UK 
             \and %inst 6 Roberto 2
            Kavli Institute for Cosmology, University of Cambridge, Madingley Road, Cambridge CB3 0HA, UK
             \and %inst 7 andy
             Department of Physics, University of Oxford, Denys Wilkinson Building, Keble Road, Oxford OX1 3RH, UK 
             \and %inst 8  %Stephane
             Sorbonne Universit\'e, CNRS, UMR 7095, Institut d'Astrophysique de Paris, 98 bis bd Arago, 75014 Paris, France 
             \and %inst 9, Matteo,  Lorenzo
             Dipartimento di Fisica e Astronomia, Universit\'{a} di Firenze, Via G. Sansone 1, 50019 Sesto F.no, Firenze, Italy
            \and %inst 10 Matteo, lorenzo and giovanni
             INAF - Osservatorio Astrofisico di Arcetri, Largo E. Fermi 5, I-50125 Firenze, Italy 
             \and   %inst 11 lorenzo 3
             University of Trento, Via Sommarive 14, I-38123 Trento, Italy
             }

   \date{\today}
 
  \abstract
  {We present a detailed study of the centre of NGC\,4654, a Milky Way-like spiral galaxy in the Virgo cluster that has been reported to host a double stellar nucleus, thus promising a rare view of ongoing star cluster infall into a galaxy nucleus. Analysing JWST NIRSpec integral-field spectroscopic data in combination with \textit{Hubble} Space Telescope Wide Field Camera 3 imaging of the inner 330 $\times$ 330 pc, we find that the NGC\,4645 nucleus is in fact more complex than previously thought, harbouring three massive star clusters within 25 pc of the centre. 
  Maps of infrared emission lines in the NIRSpec spectra show different morphologies for the ionised and molecular gas components. 
  The emission from molecular hydrogen gas is concentrated at the NSC location, while emission from hydrogen recombination lines is more extended beyond the central cluster. The velocity fields of both gas and stars indicate that the three clusters are part of a complicated dynamical system, with the NSC having an elevated velocity dispersion in line with its high stellar mass. 
  To investigate the stellar populations of the three clusters in more detail, we use surface brightness modelling to measure their fluxes from ultraviolet to mid-infrared wavelengths. This information, together with spectroscopically derived extinction values, are then used to fit the spectral energy distributions (SEDs) of the clusters. 
  Two of the clusters are UV-bright and well described by single stellar populations with young ages ($\sim$ 3 and 5 Myr) and relatively low masses ($M_\ast \sim 4 \times 10^{4} M_\sun$ and $M_\ast \sim 10^{5} M_\sun$, respectively), whereas the central cluster is much more massive ($M_\ast = 3 \times 10^7 M_\sun$), and cannot be fitted by a single stellar population. Instead, we find that the presence of a minor young population ($\sim$ 1\,Myr, $M_\ast \sim 3 \times 10^{4} M_\sun$) embedded in a dominant old population ($\sim$ 8\,Gyr) is required to explain its SED.  Given its complex composition and the close proximity of two young star clusters that are likely to merge with it within a few hundred million years, we consider the nucleus of NGC\,4654 a unique laboratory to study NSC growth from both in-situ star formation and the infall of star clusters.}

               
   \maketitle

\section{Introduction}
Nuclear star clusters (NSCs) are extremely massive and dense stellar systems \citep{Neumayer2020}, with typical stellar masses between $10^5$ and $10^8\,M_\sun$ and half-light radii of a few parsecs. They are found in at least 70\% of all galaxies, with a nucleation fraction above 80\% for galaxies in the mass range $M_\ast \sim 10^9 - 10^{10} M_\sun$ \citep{SanchezJanssen2019, Hoyer2021}. Because of their high stellar densities, NSCs are often discussed as possible sites of secular black hole formation \citep{Askar2022, Kritos2023, ArcaSedda2023}, and many NSCs are known to host a supermassive black hole (SMBH) in their centre \citep{Neumayer2020}, the Milky Way NSC being the best-studied example \citep{Schoedel2009, Schoedel2014, Gravity2019}.

The existence of tight scaling relations that connect NSC and SMBH masses to properties of the host galaxy over several orders of magnitude is generally interpreted as evidence for the co-evolution between the stellar body of the host and the growth of its central massive object \citep{Ferrarese2006, NeumayerWalcher2012, ScottGraham2013}. Consequently, understanding the formation pathways of NSCs provides important constraints on co-evolution scenarios.

In general, there are two main scenarios of NSC formation \citep[see][]{Neumayer2020}: \textit{i)} \textit{in-situ} star formation in the galaxy centre from fragmenting gas or \textit{ii)} formation through the mergers of star clusters that undergo dynamical friction and spiral to the galaxy centre. While these star clusters can be young or even gas-rich \citep{Agarwal2011, Guillard2016}, most $N$-body simulations and semi-analytical approaches have considered the gas-free mergers of globular clusters (GCs, e.g. \citealt{Tremaine1975,  CapuzzoDolcetta1993, CapuzzoDolcetta2008a, Hartmann2011, Antonini2012, Antonini2013, ArcaSedda2014}).

As GCs are typically characterised by old stellar populations ($> 10$ Gyr) and low metallicities (see \citealt{Brodie2006} or \citealt{Beasley2020} for reviews), this so-called GC-accretion channel has been invoked to explain the old, metal-poor NSCs found in dwarf galaxies \citep{Fahrion2020, Johnston2020, Fahrion2021, Fahrion2022}. Similarly, \cite{Hannah2021} found sub-dominant metal-poor populations ([M/H] $<$ -1.0 dex) in the otherwise enriched NSCs of three galaxies using spatially resolved spectroscopy, which they attributed as a contribution from the accretion of GCs. Also the Milky Way NSC was found to host a small fraction of metal-poor stars ([Fe/H] < $-0.5$ dex \citealt{FeldmeierKrause2020, Do2020, Chen2023}) possibly stemming from GCs \citep{Do2020, ArcaSedda2020}.

Nonetheless, the presence of young stars or even ongoing star formation found in many NSCs, such as the one in the Milky Way \citep{Genzel2010, Schoedel2014}, shows that gas-free accretion of GCs is not sufficient to fully explain NSC formation. Instead, NSCs might also grow \textit{in-situ} from accreted gas (e.g. \citealt{Loose1982, Bekki2006, Bekki2007, Hopkins2010, Aharon2015, Antonini2015}). The \textit{in-situ} channel naturally explains the presence of young, metal-rich populations and the complex star formation histories found in the NSCs of massive spiral galaxies \citep{Walcher2006, Kacharov2018}, as well as the super-solar metallicities found in old NSCs of massive elliptical galaxies \citep{Turner2012, Fahrion2021}.
There are a number of proposed mechanisms for funneling the gas required for the \textit{in-situ} formation into the central few pc, including gas-rich mergers \citep{MihosHernquist1994}, bar-driven gas infall \citep[e.g.][]{Schinnerer2003,Schinnerer2007}, and magneto-rational instabilities \citep{Milosavljevic2004}. On the other hand, the inspiral of young and possibly gas-rich star clusters could also explain young stellar populations in NSCs \citep{Agarwal2011, Guillard2016, ArcaSedda2016b}.

In general, it is difficult to differentiate between pure \textit{in-situ} star formation and the merger of young star clusters, as both can lead to young and metal-rich populations in NSCs.
Spatially resolving the internal kinematics of NSCs can provide additional constraints because the dissipative processes in the \textit{in-situ} formation can leave dynamical imprints such as rotating, elongated NSCs \citep{Seth2006, Brown2018, Pinna2021}. Unfortunately, this characteristic by itself is not sufficient to discriminate between the two scenarios, because under favourable conditions, the mergers of star clusters can also result in a rotating NSC \citep{PeretsMastrobuonoBattisti2014, Tsatsi2017, Lyubenova2019}. For this reason, the best evidence for NSC formation via the accretion of young star clusters comes from systems where the individual star clusters have not yet merged in the centre. Examples of this include \cite{Paudel2020} who identified off-centre young star clusters in a sample of low-mass early-type galaxies that will likely merge to build a seed NSC, as well as \cite{ArcaSedda2016b} who studied the young star clusters in the central region of the dwarf galaxy Henize 2-10, and concluded that they will likely build a NSC within a few Myr. 

The star-forming galaxy NGC\,4654 provides additional evidence for NSC formation from the accretion of star clusters, as \textit{Hubble} Space Telescope (HST) Wide Field Planetary Camera 2 (WFPC2) photometry of its centre has revealed the presence of a double nucleus \citep{GeorgievBoker2014}. \cite{Schiavi2021} further analysed this data set and built an $N$-body model that suggested that the two identified star clusters will merge within a few Myr. In this work, we present a more detailed investigation into the complex star cluster system found in the nucleus of NGC\,4654, using recent JWST NIRSpec integral-field spectroscopy (IFS) and HST Wide Field Camera 3 (WFC3) imaging. 

NGC\,4654 is a star-forming galaxy in the Virgo galaxy cluster and has been considered as a Milky Way analogue due to its spiral morphology and stellar mass of $\sim 1 - 3 \times 10^{10} M_\sun$ \citep{Lizee2021, Schiavi2021}. NGC\,4654 is likely undergoing interactions with its close neighbour NGC\,4639, as evident from tidal features and an extended CO tail indicating ram-pressure stripping \citep{Lizee2021}. The distance to NGC\,4654 is rather uncertain, with estimates ranging from 13 to 22 Mpc (see \citealt{Schiavi2021}). In this work, we adopt a value of 22 Mpc, which is the distance to the companion galaxy NGC\,4639, as determined with Cepheids \citep{Freedman2001}, following the choice of the Physics at High Angular Resolution in Nearby Galaxies (PHANGS) survey \citep{Anand2021, PHANGSHST}. At this distance, 1\arcsec\,corresponds to 107 pc. 

Our analysis is structured as follows: 
In Sect. \ref{sect:data}, we describe the JWST NIRSpec and HST WFC3 data used in this paper. In Sect. \ref{sect:spectroscopy}, we report our spectroscopic analysis of the NIRSpec data, focusing on the kinematics of stars and gas, as well as emission line ratios used to constrain the extinction towards the stellar clusters in the centre of NGC\,4654. 
In Sect. \ref{sect:imfit}, we present our surface brightness modelling of the WFC3 and NIRSpec images, used to obtain cluster sizes and fluxes across the full wavelength range from the UV to the infrared. The extinction-corrected photometry values are then used to model the spectral energy distributions (SEDs) of the star clusters, in order to derive their ages and masses, as described in Sect. \ref{sect:sed_fitting}. We discuss our results in the context of NSC formation scenarios for NGC\,4654 and other galaxies in Sect. \ref{sect:discussion}, and conclude in in Sect. \ref{sect:conclusions}.

\section{Data}
\label{sect:data}
In this work, we used JWST NIRSpec IFS observations of the nucleus of NGC\,4654 in combination with archival HST WFC3 images in five filters from the PHANGS survey. We describe the data set and our reduction methods in the following.

\subsection{JWST NIRSpec Data}
\label{sect:JWST_data}
The Near-Infrared Spectrograph \citep[NIRSpec,][]{jakobsen2022} is the prime instrument onboard JWST for spectroscopic studies in the wavelength range from 0.6 - 5.3$\,\mu$m. Our NIRSpec data were obtained as part of the JWST Guaranteed Time Observations (GTO) programme 1223 (PI: B\"oker) on 30 June 2023. The observations were acquired using the NIRSpec IFS mode described in \cite{Boker2022} that provides a 3.1\arcsec\, $\times$ 3.2\arcsec\,field-of-view. The dataset consists of four dithered exposures with a total integration time of 933.7\,s in each of the three different high resolution (R $\sim$ 2700) configurations G140H/F100LP, G235H/F170LP, and G395H/F290LP, covering the wavelength ranges 0.97 – 1.82 $\mu$m, 1.66 – 3.05 $\mu$m, and 2.87 – 5.14 $\mu$m, respectively. 

We used the STScI pipeline v1.8.2 with CRDS context 1097. A patch was included to correct some important bugs affecting this specific pipeline version (see details in \citealt{Perna2023}). We corrected the count-rate images for 1/f noise through a polynomial fit. During stage 2, we masked pixels at the edge of the slices (one pixel wide) to conservatively exclude pixels for which the correction for the throughput of the spectrograph optics (contained in the SFLAT reference files) is unreliable. We also implemented the outlier rejection of \cite{DEugenio2023}. The combination of the four dithers (with drizzle weighting) allowed us to sub-sample the detector pixels, resulting in cube spaxels with a size of 0.05\arcsec (corresponding to $\sim$5pc per spaxel).

The resulting data cubes are known to show sinusoidal modulations in single spaxel spectra, caused by the undersampling of the point-spread function (PSF). To correct these so-called ``wiggles'', we applied the methodology presented in \cite{Perna2023}. In short, the sinusoidal pattern of the brightest spaxel is fitted to establish the frequency trend, which is then used to correct the spectra extracted from neighbouring spaxels. 

In order to allow a consistent analysis of the combined HST and JWST data set, we created synthetic broad-band NIR images from the NIRSpec cubes, across passbands corresponding to a set of NIRCam filters. We chose here five different NIRCam filters and applied their transmission curves to the NIRSpec IFS cubes using \textsc{synphot} and \textsc{stsynphot} \citep{synphot, stsynphot}. We chose the NIRCam filters F115W, F162M, F200W, F335M, and F480M to avoid the wavelength gap caused by the physical separation between the two NIRSpec detectors. Applying the filter transmission curves to every spaxel in the respective cubes results in NIRCam-like images sampled at 0.05\arcsec\,pix$^{-1}$. 

\subsection{Hubble Space Telescope Data}
\label{sect:HST_data}
NGC\,4654 has been observed multiple times with HST. In 1994 and 2001, the galaxy was observed with HST WFPC2 in the filters F475W, F606W, and F814W. While the images in the F475W and F814W filters place the nucleus on the wide field camera (WFC) which only has a pixel sampling of 0.1\arcsec\,pix$^{-1}$, the F606W data place the nucleus on the planetary camera (PC) with a sampling of 0.05\arcsec\,pix$^{-1}$. Using these data, \cite{GeorgievBoker2014} reported the detection of a 'double nucleus', consisting of two star clusters. \cite{Schiavi2021} then used the same data set to further analyse the  nucleus of NGC\,4654 with structural modelling as input for a $N$-body simulation that shows that these clusters will merge within 100 Myr.

In December 2019, NGC\,4654 was observed with HST WFC3 in the UVIS channel as part of the PHANGS-HST survey \citep{PHANGSHST}. This newer data set consists of images in five filters (F275W, F336W, F438W, F555W, and F814W), covering wavelengths from the ultraviolet to the red optical regime, with higher spatial sampling than the WFPC2 data. Exposure times range from 2190\,s (F275W) to 670\,s (F555W). We downloaded the calibrated, flat-fielded and for charge transfer efficiency-corrected \textsc{.flc} exposures from the MAST archive and processed them with \textsc{DrizzlePac} \citep{DrizzlePac1, DrizzlePac2} following the standard procedure to create mosaics. The final images have a spatial sampling of 0.04\arcsec\,pix$^{-1}$ at an angular resolution of 0.067\arcsec (full width at half maximum of the PSF at 5000\AA).

\subsection{False-colour RGB images}
\begin{figure}
    \centering
    \includegraphics[width=0.48\textwidth]{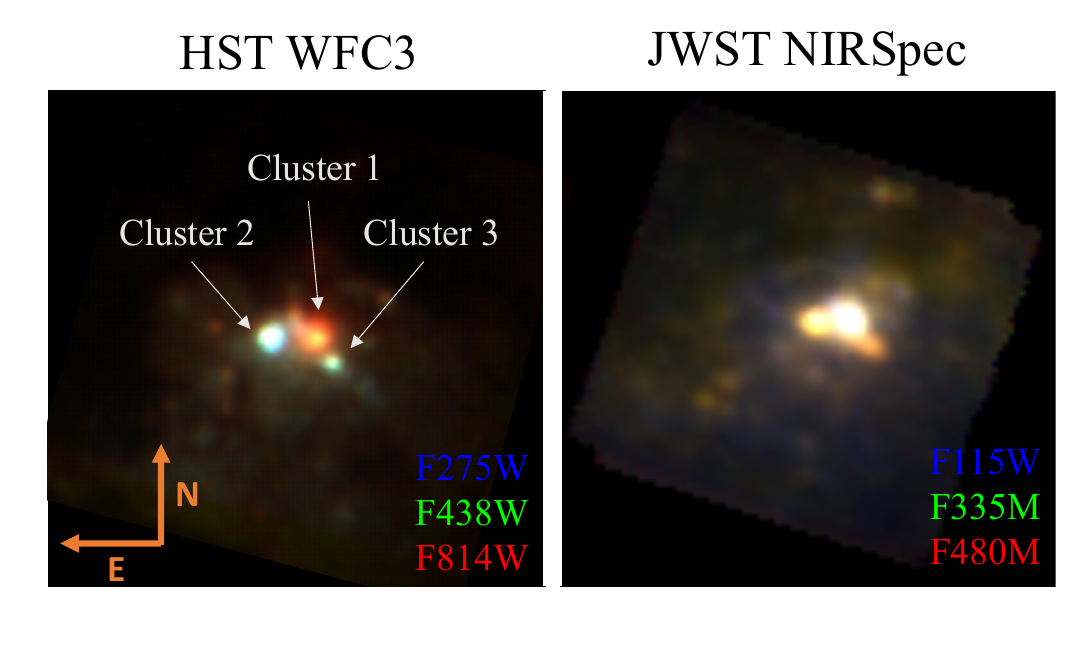}
    \caption{False-colour RGB images from HST and JWST (\textit{right}). \textit{Left}: RGB image from three HST WFC3 filters showing a 3.8\arcsec $\times$ 3.8\arcsec\,FOV. Red: F814W, green: F438W, blue: F275W. \textit{Right}: Synthetic NIRCam-like images obtained from the NIRSpec cubes showing the 3.2\arcsec $\times$ 3.1\arcsec\, FOV (340 $\times$ 330 pc). Red: F480M, green: F360M, blue: F115W.}
    \label{fig:RGB_images}
\end{figure}

Figure \ref{fig:RGB_images} shows false colour RGB images of the NGC\,4654 nucleus in log-scale. The left panel is composed from three HST WFC3 images, layered from blue (F275W) over green (F438W) to red (F814W). The right panel shows a similar image, but composed of synthetic images created from the NIRSpec data corresponding to the NIRCam filters F115W (blue), F335M (green), and F480M (red).  A number of interesting facts can be deduced directly from these images.

Both images reveal that the nucleus of NGC\,4654 not only hosts the two sources (cluster 1 and 2) identified by \cite{GeorgievBoker2014}, but at least three separate star clusters. As the blue colours in the HST RGB composite show, two of these clusters (cluster 2 and 3) are bright in the UV, indicating prominent young stellar populations. The eastern source (cluster 2) was already identified as a young star cluster by \cite{GeorgievBoker2014} and \cite{Schiavi2021}. The fainter one to the south west (cluster 3) of the centre was also found in their analysis as a fainter source labelled K3 but they were unable to measure its properties due to the limited sensitivity. As evident from its red colour in the HST image, the central cluster (cluster 1) becomes brighter at longer optical wavelengths, suggesting an older population and consequently a higher stellar mass. 
In the JWST wavelength range, the central cluster remains the brightest source at any wavelength, but the contrast to the other two is reduced towards the red end of the NIRSpec range. For this reason, the two young star clusters appear red in the JWST RGB image. 

As also discussed in \cite{GeorgievBoker2014} and \cite{Schiavi2021}, the differences in colour and brightness between the clusters indicate a difference in stellar mass. As the most massive cluster, we consider cluster 1 to be the genuine NSC of NGC\,4654 that has persisted on timescales of billions years (see Sect. \ref{sect:sed_fitting}). In contrast, the other two clusters likely are much younger objects that will merge with cluster 1 within a few hundred million years \citep{Schiavi2021}. 

In addition to the prominent three star clusters, the RGB images reveal an asymmetric dust feature in the centre of NGC\,4654, seen in the redder colours in both images towards the north. There are also two arcs near the clusters that appear blue in the JWST image and pale red in the HST image, possibly created by fainter star clusters, and a number of other faint sources. Characterising these is challenging and beyond the scope of this paper, but it is possible that they are fainter star clusters.

\section{Spectroscopic Analysis}
\label{sect:spectroscopy}

In the following, we describe the spectroscopic analysis of the NIRSpec data.
We first show the aperture spectra of the three star clusters for a qualitative comparison, obtain flux measurements of hydrogen recombination lines, and use their ratio to estimate the extinction across the NIRSpec field-of-view. This is used later in the paper to correct the photometry values in the SED modelling (Sect. \ref{sect:sed_fitting}). Additionally, we report the stellar and gaseous kinematics from the NIRSpec IFS data.

\begin{figure*}
    \centering
    \includegraphics[width=0.93\textwidth]{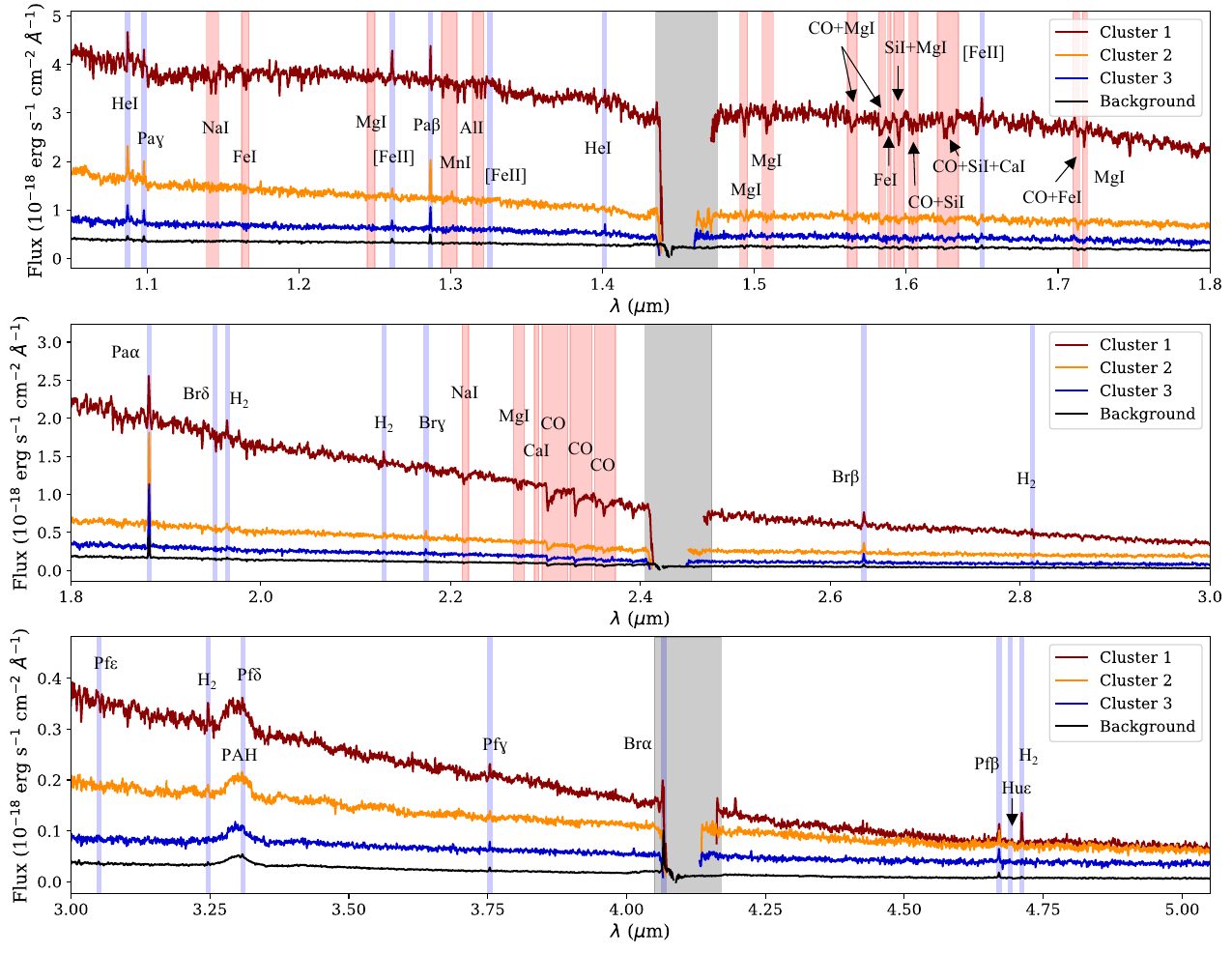}
    \caption{Spectra of the clusters in $r=0.1\arcsec$ apertures and the galaxy background annulus in the different gratings of NIRSpec. Identified emission lines are shown in blue, absorption lines in red. The grey-shaded regions mark wavelengths that are lost due to the physical separation between the two NIRSpec detectors.}
    \label{fig:cluster_spectra}
\end{figure*}

\subsection{Star cluster spectra}
We extracted the spectra of the three star clusters from the three NIRSpec data cubes using circular apertures with 2 spaxel radii (0.1\arcsec). To assess the contribution from the underlying galaxy, we further extracted a spectrum of the underlying galaxy using an annulus aperture mask with inner radius of 14 spaxels (0.7\arcsec) and outer radius of 20 spaxels (1\arcsec). 

Figure \ref{fig:cluster_spectra} shows the spectra of the three star clusters and the galaxy background in the three NIRSpec gratings. Identified emission lines are indicated by the blue shaded regions, while red shaded regions mark absorption features following the line index definitions from \cite{Riffel2019}. All spectra contain a high number of emission lines, in particular hydrogen recombination lines from the Paschen, Brackett, and Pfund series. Other emission lines include helium, iron and molecular hydrogen transitions. At 3.3 $\mu$m, the spectra show a prominent emission feature produced by polycyclic aromatic hydrocarbons (PAHs, see for example \citealt{Sandstrom2023, Lai2023}). 
Various absorption lines are also visible, especially in the high signal-to-noise (S/N) spectrum of cluster 1, where several iron, sodium, and magnesium lines can be identified. The prominent CO absorption bands between 2.3\,$\mu$m and 2.4\,$\mu$m are produced in the atmospheres of evolved stars, and indicate the presence of red giants or supergiants, especially in cluster 1. 
A more quantitative analysis of the different spectral features will be presented in future work.

Differences in the SEDs of the three clusters are already apparent from a qualitative comparison between their continuum shapes. While cluster 1 is the brightest source at short wavelengths, its flux drops fast towards long wavelengths. At $\sim$ 5 $\mu$m, the aperture spectrum of cluster 1 almost drops to the flux level of cluster 2, but it remains the brightest cluster over the full NIRSpec wavelength range. The spectrum of the underlying galaxy background shows similar features as the clusters, but at a lower flux level due to the stark contrast between the bulge population and the star clusters.

\subsection{Emission line analysis: flux maps and extinction}
\label{sect:line_fitting}
We fitted the most prominent emission lines with Gaussian profiles to derive their fluxes using \textsc{specutils}. As a first step, the spectra are corrected for Milky Way foreground extinction assuming $A_V = 0.0693$ mag\footnote{\url{https://irsa.ipac.caltech.edu/applications/DUST/}} \citep{SchlaflyFinkbeiner2011} and the extinction law of \cite{Fitzpatrick1999}. 
In order to estimate and subtract the continuum beneath each emission line, we fitted a polynomial profile over a 400 $\AA$ wide window around each line, excluding a 40 $\AA$ wide window around the line itself. The fitted continuum is subtracted from the spectrum and then the line is fitted with a Gaussian curve.
We applied this line fitting to the extracted cluster spectra, and also to every spaxel in the data cubes to create maps of emission line fluxes. We only consider fluxes that have a S/N > 3 in the respective line.

\begin{figure*}
    \centering
    \includegraphics[width=0.95\textwidth]{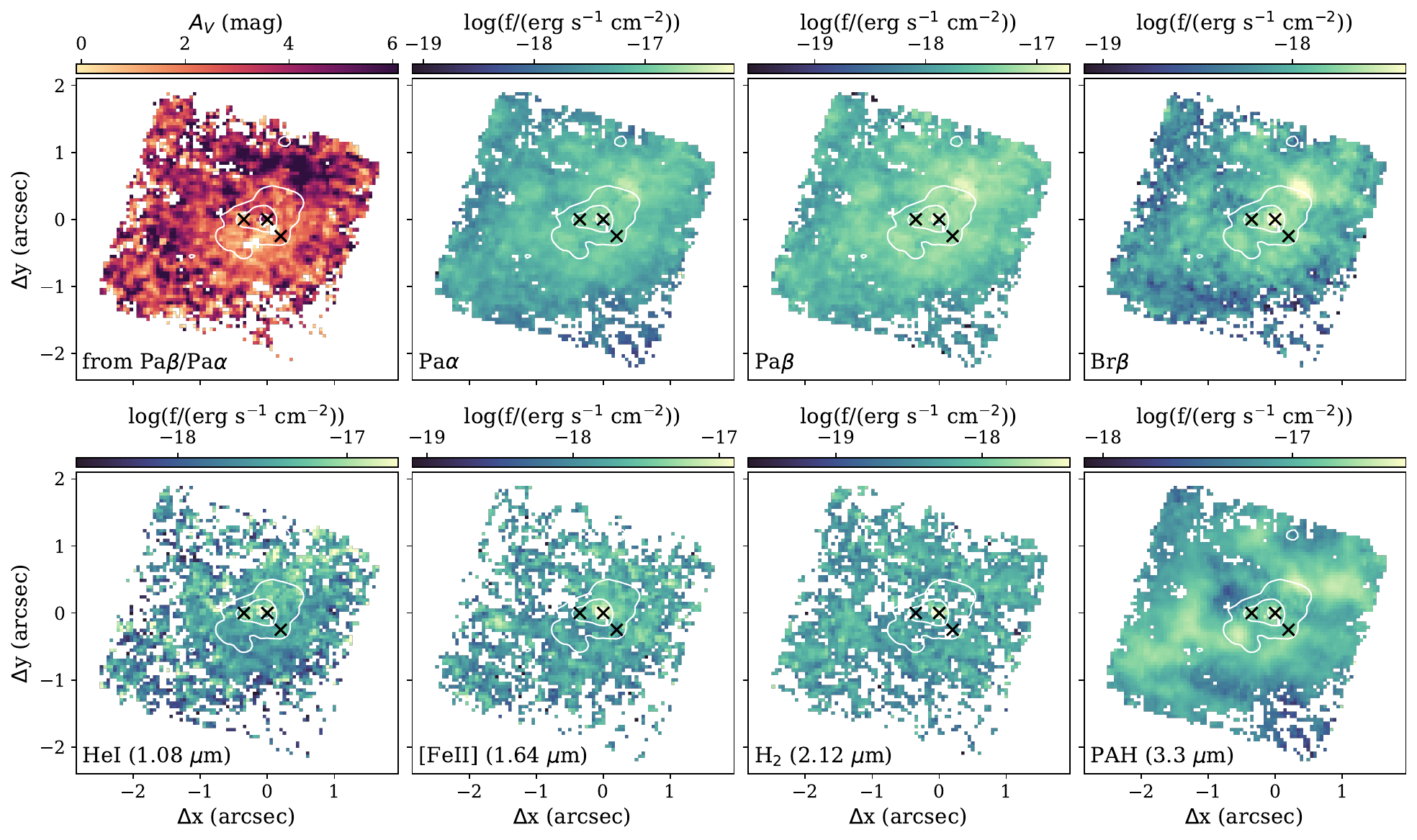}
    \caption{Maps of extinction and emission line fluxes. The top left panel shows the extinction map derived from the Paschen decrement. The remaining panels show flux maps for the Pa$\alpha$, Pa$\beta$, Br$\beta$, HeI (1.08\,$\mu$m), [FeII] (1.64\,$\mu$m), H$_2$ (2.21\,$\mu$m), and PAH (3.3\,$\mu$m) lines. These were corrected for extinction using the extinction map. Black contours show flux values from the F360M image for visualisation. Crosses mark the positions of the three clusters. Only spaxels with a S/N > 3 in the respective line are shown.}
    \label{fig:emission_line_maps}
\end{figure*}

As a first step in the subsequent analysis, we determined the dust extinction using the 
Paschen decrement (ratio of Pa$\beta$ and Pa$\alpha$), analogous to the Balmer decrement method \citep[e.g.,][]{Groves2012, Belfiore2023}. Assuming Case B recombination at a temperature of 10000 K and a density of 100 cm$^{-3}$, the intrinsic value of this ratio should be 0.48 \citep{StoreyHummer1995}. From this, the reddening $E(B-V)$ can be derived using:
\begin{equation}
    E(B-V) = \frac{\text{log}(r_\text{obs}/r_\text{0})}{-0.4 (\kappa(\lambda_1) - \kappa(\lambda_2))},
\end{equation}
where $r_\text{obs}$ is the observed flux ratio and $r_\text{0}$ is the intrinsic flux ratio of 0.48. $\kappa(\lambda)$ describes the assumed extinction law at a given wavelength, here the wavelengths of Pa$\alpha$ and Pa$\beta$. To then derive $A_V$, we assumed the \cite{Calzetti2000} extinction law for starburst galaxies with $R_V = A_V/E(V-B) = 4.05$. 

The first panel in Fig. \ref{fig:emission_line_maps} shows the extinction $A_V$ across the NIRSpec field-of-view. As can be seen, the extinction is rather low at the position of the star clusters and increases towards the north.
Applying the same method to derive the extinction from the high S/N cluster spectra directly, we find values of $A_V \sim 1.2$ mag for the three. This agrees with the maps at the respective positions, however, the maps show spaxel-to-spaxel variations.

The other panels in Fig. \ref{fig:emission_line_maps} then show the extinction-corrected emission line fluxes in the Pa$\alpha$, Pa$\beta$, Br$\beta$, HeI (1.08$\mu$m), [FeII] (1.64 $\mu$m), H$_2$ (2.21 $\mu$m), and PAH (3.3 $\mu$m) lines. The maps of the hydrogen recombination lines show the same overall structure and reveal a region of high emission towards the north west of cluster 1. This feature is also visible in the RGB images (Fig. \ref{fig:RGB_images}) as a faint arc-like structure. We note that while the derived high extinction in this region consequently boosts the extinction-corrected fluxes, this structure is also clearly visible in the uncorrected flux maps and appears as a region of high fluxes even in much redder hydrogen recombination lines (e.g. from the Brackett series). The HeI map reaches its highest values around cluster 1 and 2, while the [FeII] map appears to be peaked on cluster 1, in line with strong Fe lines seen in its spectrum. H$_2$ also peaks at this position and otherwise shows a smooth distribution that appears to roughly follow the extended distribution seen in the PAH emission. 

The concentration of H$_2$ gas at the position of cluster 1 could imply very recent or ongoing star formation at the position of cluster 1. Given that we assume it to be at the bottom of the gravitational potential well, this gas is also more likely to be retained. The presence of HeI further indicates the presence of young, massive stars, whereas the peaked [FeII] emission could indicate gas excited by shocks from recent supernovae \citep{AlonsoHerrero1997, Pasquali2011}. We leave a more detailed analysis of the chemistry in this region to future work, but these flux maps already indicate the presence of very young stellar populations in cluster 1, as will be further discussed in Sect. \ref{sect:sed_fitting}.

\subsection{Velocity fields}
\label{sect:velocities}
We measured the line-of-sight velocity and velocity dispersion of the stars by fitting the absorption lines in the spectra. In principle, all three NIRSpec cubes can be used to derive the stellar kinematics, however, as the stellar continuum drops strongly with wavelength (see Fig. \ref{fig:cluster_spectra}), and so the associated S/N, we decided to use the G140H/F100LP data. Additionally, this grating shows a high number of absorption features. In App. \ref{sect:bandheads}, we present an alternative analysis of the G235H data.
To increase the S/N, we spatially binned the data with the Voronoi binning scheme using the \textsc{VorBin} routine \citep{Cappellari2003} to achieve a suitable S/N. Binning the data to an average S/N of 50 in the continuum per wavelength bin results in 160 individual bins. 

\begin{figure}
    \centering
    \includegraphics[width=0.48\textwidth]{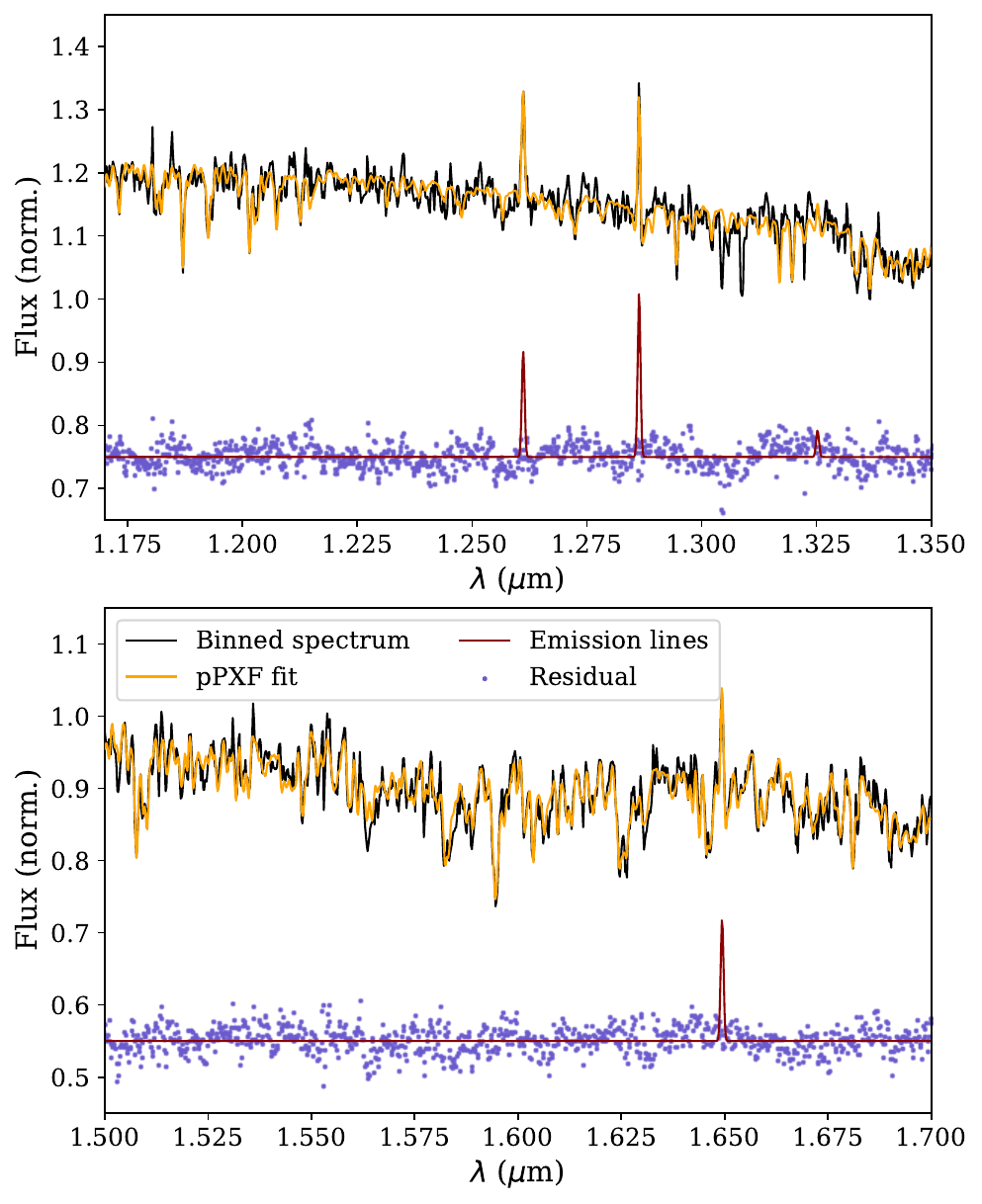}
    \caption{Example of pPXF fit to the spectrum of cluster 1 from the G140H grating. The original spectrum is shown in black. The \textsc{pPXF} fit using the XSL SSP models is shown in orange. This includes emission lines as shown in dark red. The residual is shown in purple, shifted to an arbitrary value for visualisation.}
    \label{fig:ppxf_fit_example}
\end{figure}

\begin{figure*}
    \centering
    \includegraphics[width=0.93\textwidth]{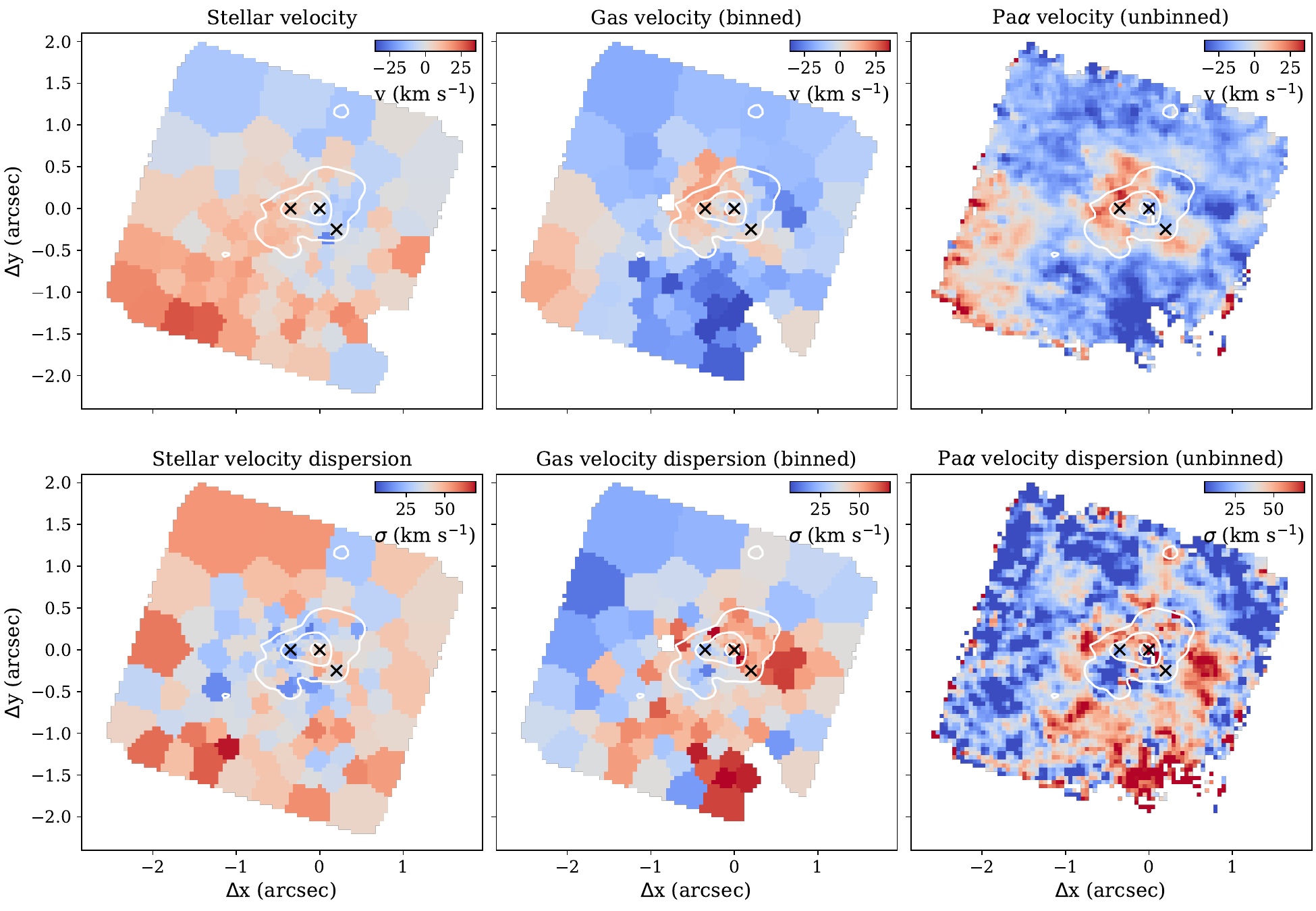}
    \caption{Kinematic maps from NIRSpec. From left to right: Maps of the stellar velocity and velocity dispersion, gas velocity from the binned data, and Pa$\alpha$ velocity from the spaxel-wise analysis. Velocities are relative to the assumed systematic velocity of 1068 km s$^{-1}$. The black contours are flux contours of the F360M image to guide the eye and the crosses mark the positions of the three clusters. Only bins with velocity uncertainties $< 10$ km s$^{-1}$ or velocity dispersion uncertainties $< 15$ km s$^{-1}$ are shown. For the unbinned Pa$\alpha$ maps, only spaxels with S/N > 10 are shown.}
    \label{fig:kinematics}
\end{figure*}

We then fitted the spectra in the individual bins using the Penalized PiXel-Fitting method (\textsc{pPXF}) \citep{Cappellari2004, Cappellari2017}, a full spectrum fitting method that fits spectra with a linear combination of user-supplied models to obtain the parameters of the line-of-sight velocity distribution such as the radial velocity and velocity dispersion. We used \textsc{pPXF} to fit the binned G140H spectra in a spectral range from 1.17 to 1.70 $\mu$m using the single stellar population (SSP) models of the X-Shooter spectral library (XSL; \citealt{XSL}). These spectra were constructed using high resolution X-Shooter spectra of stars in the Milky Way and provide a broad wavelength coverage from 0.350 to 2.480 $\mu$m, at a spectral resolution of R $\sim$ 10000, corresponding to a velocity resolution of 16\,km\,s$^{-1}$. The XSL SSP models were convolved to match the varying instrumental resolution of NIRSpec using the published values for the G140H grating
that range from $R$ = 2200 at 1.17\,$\mu$m to $R$ = 3350 at 1.70\,$\mu$m \citep{jakobsen2022}. 
We used additive polynomials of degree 12 to model the continuum. In addition, \textsc{pPXF} allows to fit emission lines with Gaussian profiles, allowing a consistency check (albeit on the binned data) with our line fitting described in Sect. \ref{sect:line_fitting}. We note that both approaches agree in the derived line fluxes, showing that our approach of modelling the underlying continuum with a polynomial agrees with the \textsc{pPXF} approach of using stellar templates.
As an example, Fig. \ref{fig:ppxf_fit_example} shows the fit to the spectrum of cluster 1. 

The resulting line-of-sight velocity and velocity dispersion maps for the stars and gas are shown in Figure \ref{fig:kinematics}. For the gas kinematics, we show the results from \textsc{pPXF} for the Voronoi-binned data and from our line fitting method on the unbinned data for the Pa$\alpha$ emission line. 
To derive reliable uncertainties for each bin, we used a Monte Carlo approach. Here, the best-fit spectrum after the first fit is randomly perturbed by Gaussian noise drawn from the residual, thereby creating a new representation of the spectrum that is then re-fitted. This is repeated 50 times for each bin and we then used the mean of the distribution as the best-fitting value, and the standard deviation as the random uncertainty. Generally, the uncertainties of the line-of sight velocity and dispersion in the individual bins are smaller than 8 km s$^{-1}$ and 15 km $^{-1}$, respectively, and only reach towards these values in the outer bins. In the plots of Fig. \ref{fig:kinematics} all bins that exceed these uncertainties are masked, which affects mainly the gas velocity and dispersion. We note here that the binned gas kinematics here refer to the \textsc{pPXF} fits with a single gas component, meaning all emission lines have the same velocity. Testing also multiple components, we found that lines other than the bright Paschen lines are associated with large uncertainties. For this reason, the gas velocity and dispersion map is mainly driven by the Pa$\beta$ line in the fitted wavelength range.

We used the radial velocity of cluster 1 (\mbox{1068 km s$^{-1}$}) as the systemic velocity, because we assume that it is located at the dynamical centre of the galaxy. This assumption is based on its high mass (see Sect. \ref{sect:sed_fitting}) which implies a very short dynamical friction time scale in case it were not at the centre (see Sect. \ref{sect:complex_system}). Given the old stellar population within cluster 1 (Sect. \ref{sect:sed_fitting}), we assume that it had ample time to sink to the bottom of the gravitational potential well. While the NIRSpec stellar kinematic map is too complex and covers too small an area of the galaxy to derive the dynamical centre directly, we note that the mean velocity within the NIRSpec field-of-view agrees with the assumed systemic velocity within the uncertainties.
% This is consistent with the mean of the velocity map (\mbox{1070 km s$^{-1}$}), given the uncertainties. Also, the NIRSpec field-of-view is not symmetric around cluster 1.

As mentioned, the stellar velocity map shows a complex structure in the centre of NGC\,4654. The three clusters are located in bins of different radial velocities. Additionally, there are indications for a large-scale rotation from the south-east to the north-west, which agrees with the global rotation in ALMA CO molecular gas maps from PHANGS-ALMA\footnote{\url{https://sites.google.com/view/phangs/home/data}} \citep{Lang2020, Leroy2021, Leroy2021b}.  We note that the kinematic centre from the CO maps coincides within the NIRSpec field-of-view, however the spatial resolution of the ALMA data is not sufficient to pinpoint the dynamical centre of the galaxy to an accuracy needed to differentiate between the three clusters.

The stellar velocity dispersion map shows an interesting structure with lower values in the central region than in the outskirts except for cluster 1 that appears as a component with high dispersion (\mbox{$\sim 45$ km s$^{-1}$}), suggesting a high stellar mass, as further quantified in Sect. \ref{fig:SED_fitting}. In contrast, the region around cluster 2 has dispersions smaller than \mbox{20 km s$^{-1}$}. We note that at this level, the velocity dispersions become unreliable due to the limited spectral resolution of NIRSpec ($\sim 30 - 40$ km s$^{-1}$). Similar velocity dispersion drops have been found in the centres of other disc galaxies (e.g. \citealt{Lyubenova2019}) and are often attributed to young stars born from dynamically cold gas (e.g. \citealt{Wozniak2003}).
The average dispersion across the map is \mbox{$\sigma = 40$ km s$^{-1}$}. While this value is lower than found in the central region of the Milky Way (\mbox{$\sigma \sim 100$ km s$^{-1}$}, \citealt{Schoedel2009, Feldmeier2014, Schultheis2021}), other spiral galaxies have also been found to have velocity dispersions \mbox{$< 50$ km s$^{-1}$} in their central regions \citep{Martinsson2013}. The ALMA CO observations of NGC\,4654 from PHANGS at 2\arcsec spatial resolution find a molecular gas dispersion of \mbox{$\sim$ 15 km s$^{-1}$} in the central region, also lower than the value of $\sim 40$ km s$^{-1}$ found in the Milky Way \citep{Schultheis2021}.

\begin{figure}
    \centering
    \includegraphics[width=0.45\textwidth]{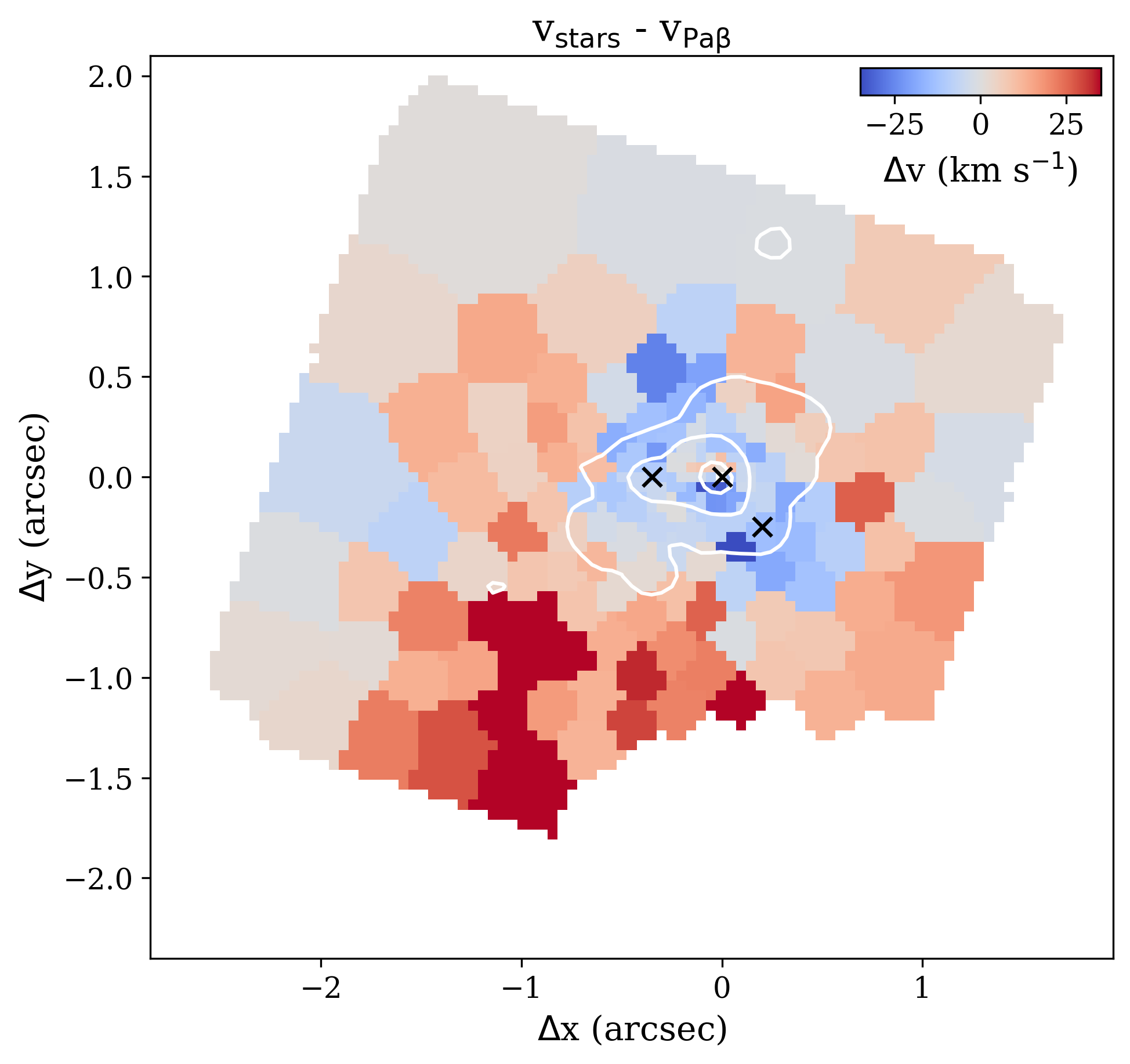}
    \caption{Difference between stellar velocity and Pa$\beta$ velocity from the binned cubes. The region where gas velocities are lower than the stellar velocities (blue colours) coincide with regions of high gas dispersions (Fig. \ref{fig:kinematics}). Only bins with velocity uncertainties $<$ 10 km s$^{-1}$ are shown.}
    \label{fig:difference_map}
\end{figure}

While the gas velocity and dispersion maps from the binned data and the Pa$\alpha$ analysis agree, they differ from the stellar kinematics in some aspects. The overall rotation of the galaxy is hinted at in the gas velocity maps, but there appears to be a structure east of cluster 1 with higher velocities. The gas velocity dispersion also appears to drop around cluster 2 and to its south. We show the difference map between stellar velocities and Pa$\beta$ (fitted here with a separate component) in Fig. \ref{fig:difference_map}. There appears a region where the gas velocities are lower than the stellar velocities around cluster 1. This region coincides with high gas velocity dispersions (Fig. \ref{fig:kinematics}), possibly indicating turbulent gas motions driven by feedback from the star formation in this central region. 

To obtain the line-of-sight velocities and dispersions of the clusters directly, we also fitted their aperture spectra with \textsc{pPXF}. In this case, we used 300 Monte Carlo iterations to establish uncertainties. Table \ref{tab:cluster_properties} reports the measured line-of-sight velocities and velocity dispersions of the clusters. 
We find differences between the cluster radial velocities in agreement with the velocity map, suggesting that they are possibly rotating around each other. While these offsets are formally not significant with respect to the uncertainties, we note that the same offsets are found when using other wavelength ranges (e.g. the G235H grating) or other SSP models (e.g. E-MILES). However, velocity uncertainties of a few km s$^{-1}$ are expected given the wavelength sampling of NIRSpec.

Additionally, and in agreement with the results from the binned maps, cluster 1 appears to have a larger velocity dispersion than the other two, which confirms that this cluster is much more massive (as also discussed in Sect. \ref{sect:sed_fitting} and Sect. \ref{sect:complex_system}). While the derived dispersions for cluster 2 and 3 are clearly well below the NIRSpec resolution, the larger dispersion of cluster 1 can reliably be measured due to its high S/N, even though it is also close to the instrumental resolution. In this context, we note that analysis of MUSE integral field spectroscopy in the recent years has shown that fitting of high S/N spectra is able to recover dispersions significantly below the instrumental resolution (e.g. \citealt{Emsellem2019}). Moreover, we found similar values for the dispersion of cluster 1 when only fitting at the red end of the G140H grating, where the resolution is below 40 km s$^{-1}$. 

We also fitted the spectra after first subtracting the background spectrum, and found velocities and dispersions that agree with the results from the unsubtracted spectra within the uncertainties, which is not surprising since the galactic background does not contribute much to the flux of cluster 1 and 2. For cluster 3, the uncertainty increases significantly when fitting the background-subtracted spectrum.

\begin{table*}
    \caption{Properties of the three clusters.}
    \centering
    \begin{tabular}{c c c c c c c c c} \hline\hline
         & R$_\text{proj}$ & v$_\ast$ & $\sigma$ & A$_{V, \text{spec}}$ & $r_\text{eff}$ & log(Age) & [M/H] & log(M$_\ast$) \\
             & (pc)    & (km s$^{-1}$) & (km s$^{-1}$) & (mag) & (pc) & log(yr) & (dex) & log(M$_\sun$) \\ 
          (1) & (2) & (3) & (4) & (5) & (6) & (7) & (8)  & (9) \\ 
                 \hline
        Cluster 1 & -- & 1068.3 $\pm$ 3.6 & 44.0 $\pm$ 4.4 & 1.20 & 9.17 $\pm$ 1.15 & 9.92$^{+0.14}_{-0.26}$ & 0.16$^{+0.15}_{-0.26}$ & 7.52$^{+0.10}_{-0.14}$ \\
        Cluster 2 & 31.8 &  1073.0 $\pm$ 4.1 & 18.8 $\pm$ 10.9 & 1.26 & 3.39 $\pm$ 0.63 & 6.71$^{+0.12}_{-0.04}$ & 0.31$^{+0.06}_{-0.20}$ & 5.10$^{+0.10}_{-0.05}$ \\
        Cluster 3 & 21.0 &  1060.0 $\pm$ 7.1 & 33.4 $\pm$ 13.6 & 1.24 & 2.52 $\pm$ 0.58 & 6.52$^{+0.10}_{-0.39}$ & -- & 4.59$^{+0.06}_{-0.06}$\\ \hline
      
    \end{tabular}
     \tablefoot{(1): Cluster identifier as in Fig. \ref{fig:RGB_images}. (2) projected distance from cluster 1. (3), (4) line-of-sight velocity and dispersion from fitting the NIRSpec spectrum. We note that the dispersions for cluster 2 \& 3 are not reliable as they are below the NIRSpec resolution. (5) extinction in the $V$-band as derived from the Paschen decrement. (6) effective radius from \textsc{imfit} modelling of the F555W image. (7), (8), and (9) mass-weighted properties of the bestfitting \textsc{bagpipes} SED models.}
    \label{tab:cluster_properties}
\end{table*}

\section{Surface brightness modelling}
\label{sect:imfit}
This section describes our analysis of the HST WFC3 and JWST NIRSpec images. We modelled the surface brightness distribution of the clusters in the F555W filter to obtain their sizes, and then derived cluster fluxes from the other images, thus covering the full wavelength range from the UV to the mid-infrared. After correction for the extinction estimated from the emission line analysis (see Sect.\,\ref{sect:spectroscopy}), these fluxes are then used to perform the SED modelling described in Sect. \ref{sect:sed_fitting} in order to estimate the ages and masses of the three clusters.

\begin{figure*}
    \centering
    \includegraphics[width=0.95\textwidth]{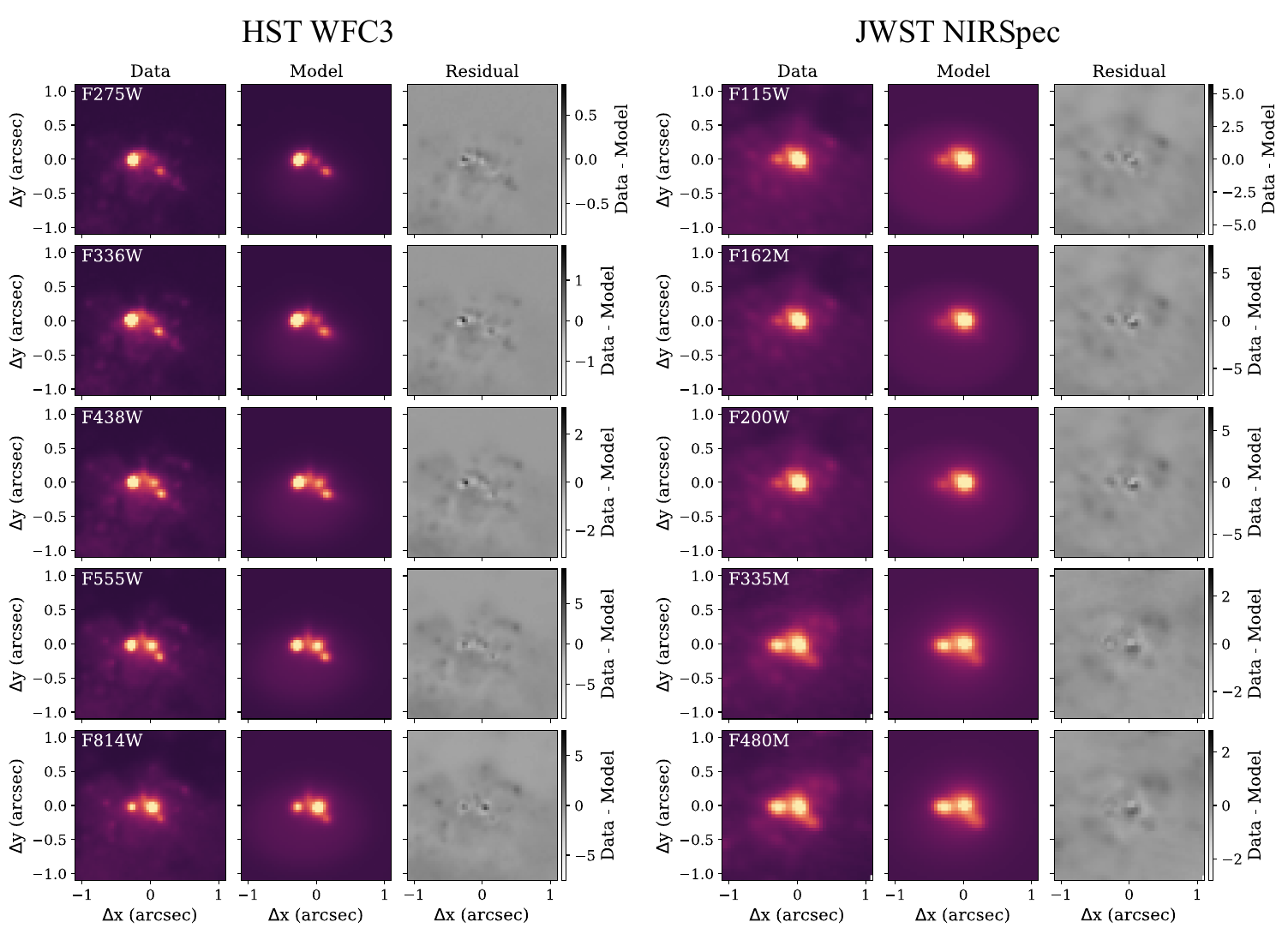}
    \caption{Surface brightness modelling in different filters. \textit{Left panels}: Cut-outs of the HST WFC3 data (\textit{left}), the imfit model (\textit{centre}) and the residual (\textit{right}). \textit{Right panels}: The same for synthetic JWST NIRCam-like images obtained from the NIRSpec cubes. The black and white extremes in the residual correspond to $\pm$5 $\sigma$, where $\sigma$ is the standard deviation of the original image in a background region. We note that there is a small difference in position angle between HST and JWST images (see Fig. \ref{fig:RGB_images}).}
    \label{fig:imfit_modelling}
\end{figure*}

\subsection{Surface brightness modelling of HST images}
We used the \textsc{imfit}\footnote{\url{https://www.mpe.mpg.de/~erwin/code/imfit/}} package \citep{Erwin2015} to build two-dimensional surface brightness models of the three star clusters, using both the HST and NIRSpec images, created as described in Sect. \ref{sect:JWST_data}. 
\textsc{imfit} is a tool for fitting various surface brightness models to galaxy images via an iterative maximum-likelihood or $\chi^{2}$ approach. After testing different combinations of models to fit the three star clusters and other components in the centre of NGC\,4654, we settled on using S\'ersic models \citep{Sersic1968} for all components. We note that using King models \citep{King1962} to describe the clusters instead of S\'ersic profiles does not appear to improve the fit quality, as they lead to similar or higher residuals. Parameter uncertainties were estimated using the implemented bootstrap functionality with 1000 iterations.

To accurately measure sizes, \textsc{imfit} convolves models with (user-supplied) instrumental PSF models. To construct reliable PSF models for the HST data, we followed the approach described in \cite{Hoyer2023} who also used \textsc{imfit} to model NSCs in external galaxies. Using the individual exposures in the \textsc{.flc} file format from the archive (calibrated exposures including charge transfer efficiency correction, but not distortion corrected), we determined the position of cluster 1 on the UVIS 1 chip of the WFC3 camera. Then we created copies of the observed exposures where all data values are set to zero to preserve the world coordinate system (WCS) information and placed \textsc{TinyTim} models \citep{Tinytim1993, Krist1995} of the PSF at the respective position. 
These PSF exposures were then combined with \textsc{DrizzlePac} in the same way as the science data (see Sect. \ref{sect:HST_data}), in order to ensure that they are sampled in the same way. We repeated this process for each of the HST filters, and used the resulting PSF images as input models for the \textsc{imfit} fitting procedure. 

We started the \textsc{imfit} modelling using the F555W HST images for a baseline, as all three clusters are well-separated in this filter. To fit the image, we first produced a 3\arcsec\,$\times$ 3\arcsec\,cutout around cluster 1 and removed the local background using the median value in the cutout. Then we constructed a five component \textsc{imfit} model, using S\'ersic profiles for each of the clusters, the slightly elongated source east of cluster 1, and a more extended component describing the contribution from the central galaxy bulge.

Each S\'ersic component has seven free parameters: the position in $x$ and $y$, position angle and ellipticity, S\'ersic parameter $n$, effective radius $r_\text{eff}$, and the intensity at effective radius $I_\text{eff}$. In the model setup for the F555W image, we left all parameters free, but provided reasonable initial guesses such as the position of the clusters and the elongated structure. 
The resulting projected distances between the clusters are listed in Table \ref{tab:cluster_properties}. We found cluster 2 to be at a distance of 31.8 pc from cluster 1, similar to the distance-corrected separations reported in \cite{GeorgievBoker2014} (0.36\,\arcsec = 38\,pc) and \cite{Schiavi2021} (0.29\,\arcsec = 31\,pc), and cluster 3 at a distance of 14.3 pc. We note that both \cite{GeorgievBoker2014} and \cite{Schiavi2021} used the WFPC2 F606W images to obtain the separation, while we used the WFC3 F555W image.
The extended ``bulge'' component does not appear to be centred on cluster 1 (or any of the other clusters), as also noted by \cite{Schiavi2021}. We suggest that this is caused by the irregular  central structure of NGC\,4654 (most likely caused by dust features) and not necessarily a true displacement of cluster 1 from the galaxy centre.

For the other WFC3 filters, we put more stringent constraints on the structural properties, but always left the intensity and radius as free parameters. This is required because the fits are less stable in the other filters, when one or more of the components become fainter (for example cluster 1 in the F275W or F335W filter). For this reason, it can be challenging to achieve reasonable fits for all five components without such restrictions. The effective radii derived from the fits to the F555W image are listed in Table \ref{tab:cluster_properties}. Figure \ref{fig:imfit_modelling} shows the resulting \textsc{imfit} models compared to the data. The morphology of the     residual images clearly illustrates that the nucleus of NGC\,4645 contains a more complex structure than described by our rather simplistic \textsc{imfit} model, but the parameters of the main components studied here are nevertheless robust. The effective radii of cluster 1 ($r_\text{eff} = 9.17 \pm 1.15$ pc) agrees within the uncertainties with the values derived by \cite{Schiavi2021} ($r_\text{eff} = 9.85$ pc), while we find a smaller radius for cluster 2 (3.4 pc rather than 6.5 pc), which might be connected to the better spatial sampling and resolution of the WFC3 data used here.

\subsection{Surface brightness modelling of NIRSpec images}

To model the NIRSpec images, we used the same five S\'ersic components. As PSF models, we used NIRSpec PSF cubes constructed from a combination of commissioning observations of a point source (PID 1128) and WebbPSF models. This is needed as the JWST PSF models available through the python based \textsc{WebbPSF} package\footnote{\url{https://webbpsf.readthedocs.io/en/latest/intro.html
}} \citep{Perrin2014} do not account for the wavefront error of the NIRSpec IFU optical path. The details of how these PSF cubes were constructed will be described in a forthcoming paper (Beck at al. in prep.). We applied the same NIRCam filter curves as described in Sect. \ref{sect:JWST_data} to the PSF cubes to obtain PSF model images that were then used in \textsc{imfit}.
Additionally, we note that we use the NIRSpec images only to derive integrated NIR fluxes of the clusters, which are rather insensitive to the exact description of the PSF. Even modelling without PSF model gives integrated flux levels that agree within the uncertainties.

As the right panel in Fig. \ref{fig:imfit_modelling} shows, the appearance of the cluster ensemble in the nucleus of NGC\,4654 changes dramatically across the NIRSpec wavelength range, creating some challenges for the \textsc{imfit} modelling. While the three clusters can be easily separated at the longest wavelengths, cluster 1 outshines the others at shorter wavelengths and particularly cluster 3 becomes barely detectable at wavelengths $\sim$ 1 - 2.5 $\mu$m. We nonetheless forced a fit at these wavelengths by keeping the component fixed in the model (except for intensity and radius), but found that the bootstrapping often prefers solutions with zero intensity. For this reason, we did not use the cluster 3 fluxes in the F115W, F162M, and F200W bands in our analysis.

\section{SED modelling with Bagpipes}
\label{sect:sed_fitting}

\begin{figure}
    \centering
    \includegraphics[width=0.48\textwidth]{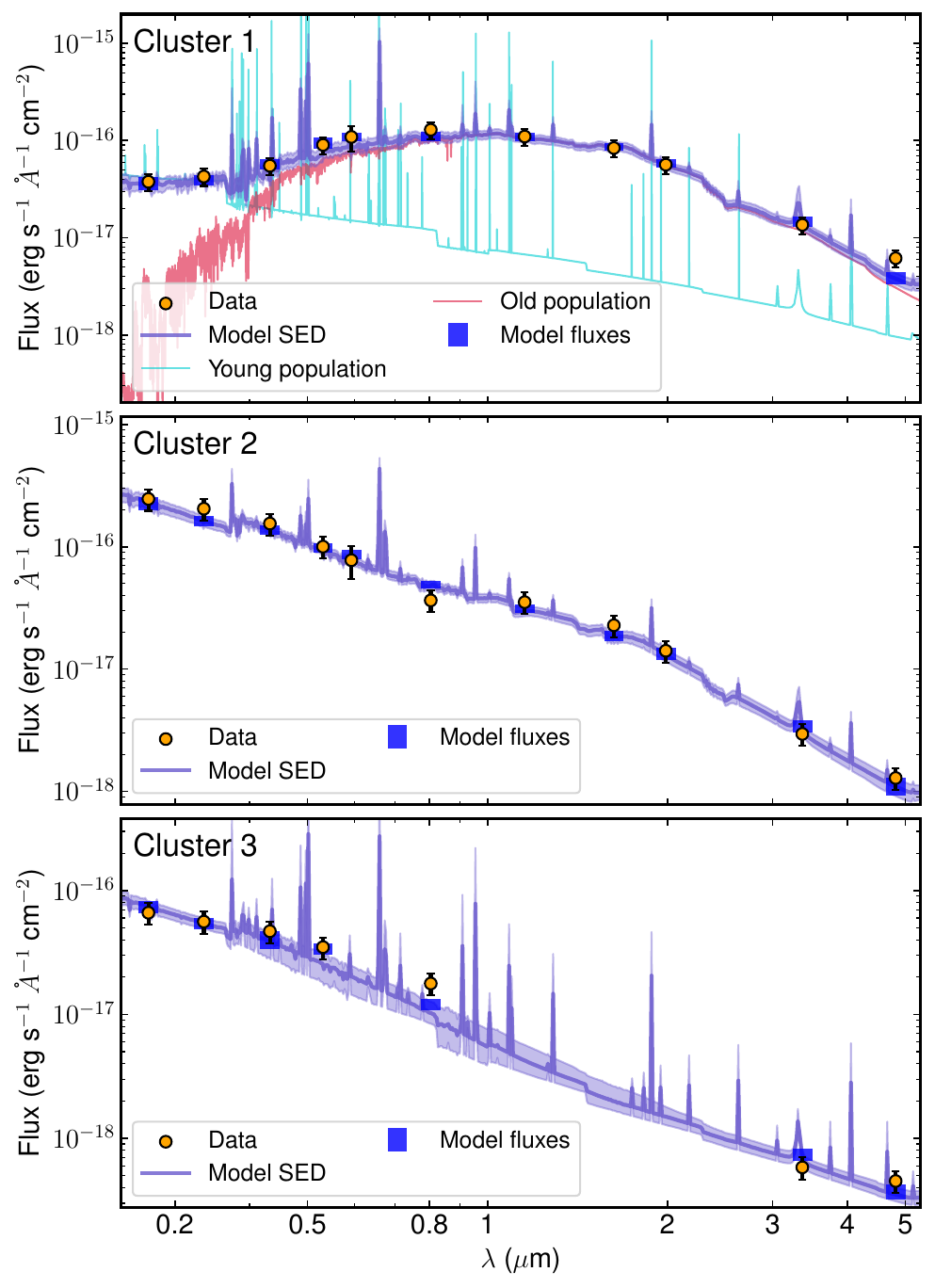}
    \caption{Results of the \textsc{bagpipes} SED fitting for the three clusters. The fluxes derived from \textsc{imfit} modelling are plotted as orange points. The best-fitting \textsc{bagpipes} SED models are shown in purple, with the shaded regions indicating the 16th and 84th percentiles of the posterior samples. The corresponding model fluxes are shown in blue. In the top panel, the young (cyan) and old (red) populations that make up the best-fitting SED for cluster 1 are shown separately. Clusters 2 and 3, in contrast, are well fitted with a single-age stellar population.}
    \label{fig:SED_fitting}
\end{figure}

To constrain the stellar population properties such as ages and metallicities of the three clusters, one could use full spectrum fitting with \textsc{pPXF} or other tools to fit model templates to the NIRSpec data. While this technique has been applied successfully to optical spectra of star clusters (e.g. \citealt{Fahrion2021, Fahrion2022}), we decided to instead use SED modelling based on the \textsc{imfit} fluxes for several reasons. Firstly, many SSP models are uncertain at young ages (e.g. \citealt{Vazdekis2016, XSL}) and often do not include ages younger than a few tens of million years, especially not for the infrared part of the spectrum. Secondly, SSP models are still largely untested in the infrared. As one example, \cite{Eftekhari2022} reported strong differences between observed and predicted CO absorption from the E-MILES models in massive elliptical galaxies, and star forming galaxies are even less well explored. Finally, SED modelling allows us to also use the HST images that cover the UV, which is essential to constrain the ages of young populations (e.g. \citealt{Adamo2017}). For these reasons, we do not pursue the full spectral modelling further in this work, but instead will use a subsequent paper to focus on a more detailed investigation of the complex chemistry in this region using both absorption and emission features.

We used the Bayesian analysis of galaxies for physical inference and parameter estimation code \textsc{(bagpipes}; \citealt{Carnall2018, Carnall2019})\footnote{\url{https://bagpipes.readthedocs.io/en/latest/index.html}} to model the SEDs of the three clusters. \textsc{bagpipes} allows the user to build flexible models to constrain the emission and absorption from stars, gas, and dust. The included stellar population models are the 2016 version of the \cite{BruzualCharlot2003}  models with a Kroupa initial mass function (\citealt{Kroupa2001}, considering stellar masses from 0.1 to 100 $M_\sun$) that provide spectra with a 2.5\,\AA\ resolution in a wavelength range from 3525 to 7500\,\AA, based on the MILES stellar spectral library \citep{FalconBarroso2011}. Due to this limited spectral range, we cannot use \textsc{bagpipes} to model the NIRSpec spectra and instead only focus on the photometry.
Nebular emission is modelled through the \textsc{CLOUDY} photoionisation code \citep{Ferland2017}, and \textsc{bagpipes} offers different prescriptions for the dust attenuation. For dust emission, energy balance is assumed and the models from \cite{DraineLi2007} are used. Extinction is assumed to be the same for stars and gas.

We modelled the SEDs using the HST WFC3 and JWST NIRSpec magnitudes derived from \textsc{imfit}, and for clusters 1 and 2, we also included the F606W magnitude measured from the older WFPC2 data by \cite{Schiavi2021}. The bootstrap method of \textsc{imfit} allows us to estimate the uncertainties of the model fluxes, and we find them to be on the order of 5\% to 15\%. However, for the SED modelling, we assumed a more conservative flux uncertainty of 20\% to also cover the variations obtained when using different parameter settings with \textsc{imfit} and the NIRSpec flux calibration and used 30\% for the F606W magnitudes from \cite{Schiavi2021} as they did not list uncertainties. We adopted these conservative uncertainties because, as described above, the five component \textsc{imfit} model is sometimes unstable in its solution, especially when some of the clusters are faint. We show the \textsc{imfit}-drived photometry of the three clusters, together with their best-fitting model SEDs in Fig. \ref{fig:SED_fitting}.

Several choices for the star formation history are available in \textsc{bagpipes}. As star clusters are generally well described by a single stellar population with a well defined age and metallicity, we first attempted to fit the cluster spectra with a single delta-peak burst for the star formation history. This burst model has age, metallicity, and the stellar mass formed in the burst as free parameters. This originally formed stellar mass evolves with the age of the star cluster due to stellar evolution to the stellar mass observed today.
Additionally, we included a component describing the attenuation from dust and its re-emission at long wavelengths. As we have measured the extinction from the spectroscopy, we applied a Gaussian prior on the extinction with mean $A_V = 1.2$ mag and a conservative standard deviation of 0.2 mag and used again the \cite{Calzetti2000} extinction law that is implemented in \textsc{bagpipes}. For the dust emission parameters, we left the PAH mass fraction free as this influences the flux in the F335M band. Lastly, we included possible nebular emission using a fixed ionisation parameter of log($U$) = $-$2.

We found that such a single burst model is able to fit the SED of cluster 2 well. The model finds a best-fitting age of \mbox{$\sim$ 5 Myr} and an mass of \mbox{10$^5$ $M_\sun$}, confirming that this cluster is very young, as expected from its UV flux. The mass-weighted values are reported in Table \ref{tab:cluster_properties} and we show the corner plots of the posterior samples in the Appendix \ref{app:corner_plots}.

In contrast to cluster 2, the SED of cluster 1 cannot be fitted by a single age population. Instead, at least two populations of very different ages are required to simultaneously describe its UV flux and near-infrared continuum slope. For this reason, we fitted the SED of cluster 1 with a more complex model containing two populations of different age, metallicity, and mass, but with a common extinction and dust emission. To constrain the extinction, we again applied a Gaussian prior based on the spectroscopic measurement (Sect. \ref{sect:line_fitting}). 
The best fit is achieved by a combination of a very young (\mbox{$\sim$ 1\,Myr}) population with a mass of \mbox{10$^{4.5}\,M_\sun$} within a dominant, old population (\mbox{$\sim$ 9 Gyr}) with a stellar mass of \mbox{$M_\ast = 3 \times 10^{7} M_\sun$}.
In the top panel of Fig. \ref{fig:SED_fitting}, we plot the resulting model SED, as well as the two individual components of cluster 1. Given that this composite model has seven free parameters fitted to only 11 flux values, some parameters of the two populations, in particular their metallicities and the age of the young population, are not well constrained as can be seen in the corner plots in App. \ref{app:corner_plots}. To better constrain this aspect, detailed spectroscopy in the optical range or with suitable NIR stellar population models are needed.

As an alternative to the model with two bursts, we also tested a model with an exponentially declining star formation history (star formation rate SFR $\propto \text{exp}(-t/\tau)$) to model the SED of cluster 1. This model has fewer free parameters, because it assumes the same metallicity for all stars. The fit finds an age of the oldest stars of $\sim$ 9 Gyr, with a $\tau = 6.4$ Gyr, illustrating again that young stellar ages are needed to fit the measured UV fluxes.  We therefore conclude that cluster 1 does not contain a single stellar population but rather consists of populations of varying ages. In particular, a minor young population is needed to explain the UV fluxes.

For cluster 3, we took a similar approach as for cluster 2, since this object is also brightest at UV wavelengths. However, this cluster is overall much fainter, and barely discernible in the short wavelength JWST passbands (F115W, F162M, and F200W), leading to large uncertainties in the measured fluxes at these wavelengths. For this reason, we only fitted the wavelengths with reliable flux measurements, and set the metallicity to that of cluster 2 to limit the number of free parameters. The best-fitting model finds this star cluster to also be very young (\mbox{Age $\sim$ 3 Myr}), but to have a lower mass than cluster 2 (\mbox{$M_\ast = 4 \times 10^4 M_\sun$}).

\section{Discussion}
\label{sect:discussion}
In the following, we discuss our results in the context of the formation history and future evolution of the NGC\,4654 nucleus, and compare our findings to other galactic nuclei. 

\subsection{The complex star cluster system in the centre of NGC\,4654}
\label{sect:complex_system}
While originally suggested to host a double nucleus, the HST and JWST data presented here clearly show that the centre of NGC\,4654 hosts a system of at least three star clusters, superposed on a rather complex stellar and gaseous background structure. As already evident from the RGB images created from the HST WFC3 and JWST NIRSpec data, our data demonstrate that two of these star clusters are significantly younger than the central cluster (cluster 1).
We also provide an updated estimate for the mass of cluster 1: while both \cite{GeorgievBoker2014} and \cite{Schiavi2021} concluded (based on HST WFPC2 data) that cluster 1 has a mass of around 10$^{6} M_\sun$, our SED modelling that also includes near-infrared JWST NIRSpec data has shown that the bulk of its stellar population is very old, resulting in a high mass-to-light ratio and consequently a much higher stellar mass of $M_\ast = 3 \times 10^7 M_\sun$. 

To test how well this stellar mass compares with a dynamical mass estimate, we can use the structural properties and measured velocity dispersion to estimate the virial mass following the estimator introduced in \cite{Spitzer1987}: \mbox{$M_\text{vir} \approx 9.75 r_\text{eff} \sigma^{2} / G$}, where $r_\text{eff}$ is the projected effective radius and $\sigma$ is the velocity dispersion. Using the values reported in Table \ref{tab:cluster_properties}, this leads to an estimated virial mass of \mbox{4.0 $\times$ 10$^{7}\,M_\sun$}, in close agreement with the inferred stellar mass from the SED modelling. 
Based on its high mass, we suggest that cluster 1 is likely the NSC proper of NGC\,4654, located at the dynamical centre of the galaxy, even though the complex kinematics in the NIRSpec data do not allow us to infer the dynamical centre directly. This assumption is supported by a simple consideration of the dynamical friction timescales \citep{BinneyTremaine1987}: a cluster as massive as cluster 1 will spiral to the centre of the galaxy within 100\,Myr, even if we assume an extreme initial separation of 1\,kpc. As this inspiral timescale is much shorter than the stellar age ($\sim$ 8 Gyr), we can conclude that cluster 1 has long been at the centre of the galaxy. In contrast, the much younger ages and lower masses of clusters 2 and 3 likely indicate that they have formed close to the NSC and will inspiral in the future.

Both \cite{GeorgievBoker2014} and \cite{Schiavi2021} have estimated the time after which cluster 2 will likely merge with the NSC based on their mass estimates, and found time scales shorter than a few hundred million years. In particular, \cite{Schiavi2021} presented a kinematic model testing different configurations and found that the merger timescale can be as short as a few million years depending on the exact orbit. Given that cluster 3 now complicates the dynamics of the system significantly, we leave a more detailed kinematic study of this system for future work, but note that both clusters likely will merge into the NSC in the near future, further contributing to its mass growth.

\subsection{The assembly of NGC 4654's nuclear star cluster}
The formation history of NSCs is a long debated topic that revolves around two main ideas: assembly from merging star clusters, which are often considered to be ancient GCs (e.g. \citealt{Tremaine1975, CapuzzoDolcetta1993}), or \textit{in-situ} star formation from newly accreted gas (e.g. \citealt{Bekki2006}). While information about their formation history usually has to be obtained from the current state of NSCs, NGC\,4654 offers a unique opportunity to study the continuous growth of its NSC in great detail.

As described above, the close projected distances of cluster 2 and 3 and small relative radial velocities mean that they will likely merge with the NSC (cluster 1) on a timescale of a few million years. Given their low masses, this will not significantly increase the mass of the NSC, but will shape the complicated stellar population properties of this cluster further. As our SED modelling showed, there are already very young stars (\mbox{$\sim$ 1 Myr}) in the NSC that are contributing significantly to its UV emission, while most of its mass stems from old populations. 
Consequently, it appears we are currently observing a hybrid formation scenario for the growth of this NSC: the accretion of fairly low-mass young star clusters (as described for example in \citealt{Agarwal2011} or \citealt{Guillard2016}), onto a NSC that is dominated in mass by populations as old as typical GCs. These old populations might have been formed directly in the galaxy centre many billion years ago, or could have been brought in with massive GCs formed at larger distances. With our limited view of the internal stellar populations we cannot discern either scenario. A detailed investigation of the metallicity in the NSC and the galaxy itself would be required, for example to identify a metallicity contrast between a metal-poor NSC and a metal-rich galaxy centre that could indicate the early accretion of metal-poor GCs from larger radii (e.g. \citealt{Fahrion2020, Johnston2020, Neumayer2020, Fahrion2021}). 

Our analysis lets us draw some conclusions about whether the young population in the NSC has formed there directly or is the result of a recently merged young star cluster. The SED modelling favours an even younger age than that of clusters 2 and 3, which might indicate that this population has truly formed within the NSC or at least at very close distance with a rapid accretion, but the uncertainties on the age are too large for a definitive conclusion based on this argument. However, further evidence for in-situ formation comes from the peaked distribution of various emission lines indicative of ongoing star formation, in particular H$_2$, HeI, and [FeII]. The presence of hot molecular gas in the NSC is also evidence for active star formation possibly trigged by gas accretion onto the NSC, in the sense that supernovae have not yet completely destroyed the natal gas clouds. The extremely young age of $\sim$ 1 Myr inferred from SED fitting is certainly in line with this scenario, as simulations of star cluster formation typically find that feedback from massive stars destroys the molecular clouds on slightly longer timescales ($\sim$ 5 Myr, \citealt{Howard2018, Krause2020, Guszejnov2022}). Given that the H$_2$ does not peak on the other two clusters is also in line with their somewhat higher ages found by the SED fitting. Lastly, the elongated structure next to cluster 1 could stem from ongoing accretion of gas or stars onto cluster 1, but a more detailed kinematic and chemical analysis is needed to confirm this.

\subsection{Comparison with other galaxies}
In the recent years, investigations of the stellar populations in NSCs compared to their host galaxies have started to paint an increasingly coherent picture of NSC formation as a process that depends on a galaxy's individual formation history. While the low-mass, old, and metal-poor NSCs in many dwarf galaxies are very similar to the GCs that might have formed them in the first place (e.g. \citealt{Fahrion2020, Johnston2020, Neumayer2020}), the more massive NSCs in massive galaxies ($M_\text{gal} > 10^{9} M_\sun$) are often found to be more complex. Explaining their formation often requires formation from either in-situ star formation directly or accretion of young, metal-rich star clusters to account for their often extended star formation histories \citep{Kacharov2018, Fahrion2021, Hannah2021}. 

In that regard, the NSC of NGC\,4654 fits well among other late-type galaxies, as we have also found evidence for multiple populations. Nonetheless, the properties of NSCs vary strongly even among late-type spirals. For example, using HST WFC3 and JWST NIRCam imaging of NGC\,628, \cite{Hoyer2023_NGC628} found its NSC to have a similarly high mass as in NGC\,4654 (\mbox{$\sim 10^7 M_\sun$}), but no evidence for recent star formation. Similarly, among the three NSCs investigated by \cite{Hannah2021} with spatially resolved spectroscopy, the NSC in NGC\,205 is younger than 1\,Gyr, while the other two are dominated by older populations. This variety in the observed stellar population ages is most easily interpreted as evidence for recurring in-situ star formation in (gas-rich) galactic nuclei, which is `sampled' at random phases by observations.  

Given its morphology and stellar mass, NGC\,4654 is often considered a Milky Way analogue, and our analysis has shown that this analogy between the two galaxies can be extended to their NSCs, as both have a mass of \mbox{$\sim 3 \times 10^{7}\,M_\sun$} \citep{Feldmeier2014, Chatzopoulos2015, }. In addition, the Milky Way NSC also contains populations of different ages and is dominated in mass by an old population (e.g. \citealt{Pfuhl2011, Schoedel2020, Chen2023}). In addition, individual stars within the Milky Way NSC likely are as young as a few million years \citep{Genzel2010, Lu2013, Schoedel2020}, similar to the 1\,Myr old population that we have identified in NGC\,4654's NSC. The presence of the two young clusters near the NGC\,4654's NSC is also reminiscent of the Milky Way, where the Arches and Quintuple clusters, two young (\mbox{2 - 5 Myr}) and massive (\mbox{> 10$^4\,M_\sun$}) star clusters \citep{Figer1999a, Figer1999b, Figer2002, Schneider2014, Clark2018}, are found at distances of 20 - 30 pc from the Galactic Centre.

In addition to the Milky Way and NGC\,4654, other galaxies are also known to host multiple star clusters in their centre. Prominently, the nucleus of M\,31 is a complex system with a bimodal distribution of stars and blue subclusters \citep{Lauer1993, Lauer2012} in the NSC, which is often interpreted as evidence for interactions between the stars on disk-orbits and the central black hole \citep{Peiris2003, Kazandjian2013}. 
For example, the starbust galaxy Henize 2-10 hosts in its centre young massive clusters that are expected to merge within few hundred million years \citep{ArcaSedda2015}, and might contain a central massive black hole (\citealt{Riffel2020}, but see \citealt{Cresci2017} and \citealt{Hebbar2019}). 

Based on HST imaging, other galaxies have also been found to have complex stellar structures with multiple brightness peaks in their centres, for example NGC\,4486B \citep{Lauer1996}, VCC\,128 \citep{Debattista2006} or IC\,676 \citep{Zhou2020}. These structures are referred to as dual nuclei or binary nuclei, a term that is also used to describe two galaxy nuclei in close separation following the not fully concluded merger of two galaxies \citep{DeRosa2019, Perna2023b}. 
For example, kinematic modelling of two nuclei in NGC\,7727 has found that this galaxy hosts two bona-fide galaxy nuclei that both host supermassive black holes \citep{Voggel2022}. 
In the future, JWST's capabilities of peering behind the dust of strongly interacting systems will further advance the study of such double galaxy nuclei (Ceci et al. in prep., Ulivi et al. in prep), but we note here that the case of NGC\,4654 shows that already a single galaxy nucleus can host   multiple star clusters of strikingly different properties.

\subsection{A massive black hole in cluster 1?}
\label{sect:SMBH?}
The coexistence of SMBHs and NSCs is well established, demonstrated already by the Milky Way NSC which hosts an SMBH with a mass of $4 \times 10^6 M_\sun$, and many other known examples (see \citealt{Neumayer2020}). Additionally, studies of the black hole occupation fraction using observations of active galactic nuclei (AGN; e.g. \citealt{Greene2012, Miller2015}) imply that the majority of galaxies in the mass range of NGC\,4654 ($M_\text{gal} = 1 - 3 \times 10^{10} M_\sun$; \citealt{Lizee2021, Schiavi2021}) host accreting black holes. Considering `quiescent' black holes that are more difficult to observe, this fraction likely rises above 70\%, as also suggested by hydrodynamical simulations \citep{Tremmel2023}. It is therefore natural to ask whether NGC\,4564's NSC hosts an SMBH, and whether our data can provide evidence to answer this question.

Going back to the cluster spectra in Fig. \ref{fig:cluster_spectra}, it is evident that coronal lines  from highly ionised atoms (e.g. [FeVI], [FeXIII], [SiVI], [SiXI], or [MgIV]; see \citealt{Cann2018}) that are often produced by an accreting black hole are absent from the spectrum. Moreover, the detected recombination lines from ionised atoms such as hydrogen (Paschen and Brackett recombination lines) or helium (HeI) do not seem to show a prominent broad component that could be associated with the Broad Line Region (BLR) of an AGN. However, the lack of highly ionised line transitions and BLR features do not definitely indicate the absence of a SMBH in the NSC (e.g. \citealt{RodriguezArdila2011, Lamperti2017, denBrok2022, Perna2024}).

Comparing the derived stellar mass of cluster 1 from the SED fitting ($\sim$ 3.3 $\times$ 10$^{7} M_\sun$) to the estimated dynamical mass ($\sim$ 4.0 $\times 10^7 M_\sun$) also provides no strong indication that there is an SMBH that would contribute to the dynamical mass, but not the stellar mass, especially given the associated uncertainties of a least 0.1 dex. Nonetheless, there is still room for even an SMBH with a mass of a few million solar masses, similar to the Milky Way NSC.

\section{Conclusions}
\label{sect:conclusions}
We have presented a spectroscopic and photometric analysis of the centre of the Milky Way-like galaxy NGC\,4654 that hosts a massive NSC surrounded by two younger star clusters. We obtained JWST NIRSpec IFS data to analyse gas emission and the kinematics of stars and gas in the centre. After creating images from the NIRSpec cubes in various NIR passbands, and combining them with multiband photometry from HST WFC3, we performed SED modelling of the clusters. We summarise our results in the following.

\begin{itemize}
    \item False-colour RGB images created from the HST and JWST data reveal stark differences in the stellar population ages between the central cluster (cluster 1) and the other two which are much brighter in the UV wavelengths, indicating young populations. In contrast, cluster 1 is brightest at longer wavelengths, indicating an older stellar population and higher stellar mass.
    \item JWST NIRSpec data from 1 to 5 $\mu$m shows a wealth of emission and absorption lines. We were able to identify many hydrogen recombination lines and used the Paschen decrement to map the extinction across the NIRSpec IFU field-of-view, finding it to increase towards the north of the central cluster. Additionally, we found the hydrogen recombination line fluxes to be distributed smoothly, while emission from excited H$_2$ is peaked on the location of cluster 1. The dust distribution traced by the 3.3\,$\mu$m PAH feature shows an extended, complex morphology.
    \item We inferred the line-of-sight velocities and velocity dispersions by fitting the NIRSpec spectra with stellar models and found a complex velocity structure of the three star clusters. The three clusters have small relative velocities. Cluster 1 clearly has the highest velocity dispersion (\mbox{$\sigma = 44$ km\,s$^{-1}$}), further evidence for its high stellar mass. A simple estimation of its dynamical mass finds it at 2.5 $\times 10^{7} M_\sun$.
    \item We modelled the two-dimensional surface brightness profiles of the clusters in the HST WFC3 and NIRSpec images using \textsc{imfit}. To account for all the structures in the centre of NGC\,4654, a complex model with five separate components is required. We found that the relative intensity between the clusters strongly changes with wavelength, again indicating differences in their ages. From the modelling, we found that cluster 2 has a projected distance of 21.6 pc to cluster 1 and cluster 3 is at a distance of 14.3 pc from cluster 1.
    \item To infer stellar population properties of the clusters, we fitted their SEDs from 0.2 to 5\,$\mu$m. Cluster 2 and 3 are well fitted with single stellar populations of very young ages ($\sim$ 5 and 3\,Myr, respectively) and low masses ($M_\ast \sim 10^5$ and $4 \times 10^4 M_\sun$), but cluster 1 required a fit with a composite population. To simultaneously reproduce the UV fluxes and the declining slope in the infrared, a very young population ($\sim$ 1\,Myr) embedded within a dominant old population (8\,Gyr) is required. We found cluster 1 to have a stellar mass of \mbox{$M_\ast \sim 3 \times 10^7\,M_\sun$}, similar to the mass of the Milky Way NSC and in agreement with the dynamical mass derived from the velocity dispersion.
    \item Given its significantly higher mass, we conclude that cluster 1 is the bona fide NSC of NGC\,4565, and that it will likely continue to grow in through impending  mergers with clusters 2 and 3. 
    The two populations found in cluster 1, combined with the presence of the two young star clusters illustrate the highly complex process of NSC formation, into which NGC\,4654 offers a unique window.
    \item We have discussed that our data provides scant, if any, evidence for the presence of an SMBH in cluster 1. Based on our analysis, we cannot rule out an SMBH with $<10^5\,M_\sun$ which is the expected black hole mass if NGC\,4654 followed the $M_\text{BH}$ - $\sigma$ relation for late-type galaxies.
\end{itemize}

\begin{acknowledgements}
We thank the anonymous referee for insightful suggestions that have helped to polish this paper.
We thank Nils Hoyer for valuable advise about imfit modelling and Ciar\`an Rogers for discussions on how to derive the extinction.
KF acknowledges support through the ESA research fellowship programme. 
MP and SA acknowledge support from the research project PID2021-127718NB-I00 of the Spanish Ministry of Science and Innovation/State Agency of Research (MCIN/AEI/10.13039/501100011033). 
AJB acknowledges funding from the “FirstGalaxies” Advanced Grant from the European Research Council (ERC) under the European Union’s Horizon 2020 research and innovation program (Grant agreement No. 789056).
GC acknowledges the support of the INAF Large Grant 2022 "The metal circle: a new sharp view of the baryon cycle up to Cosmic Dawn with the latest generation IFU facilities".
Based on observations with the NASA/ESA {\em Hubble Space Telescope} and the NASA/ESA/CSA {\em James Webb Space Telescope}, which are operated by AURA, Inc., under NASA contracts NAS5-26555 and NAS 5-03127. This work made use of Astropy:\footnote{\url{http://www.astropy.org}} a community-developed core Python package and an ecosystem of tools and resources for astronomy \citep{astropy2013, astropy2018, astropy2022}.
\end{acknowledgements}

\bibliographystyle{aa} % style aa.bst
\bibliography{references}

\begin{appendix}
\onecolumn
\section{Fitting the CO bandheads at 2.3 $\mu$m}
\label{sect:bandheads}
\begin{figure*}
\centering
    \includegraphics[width=0.90\textwidth]{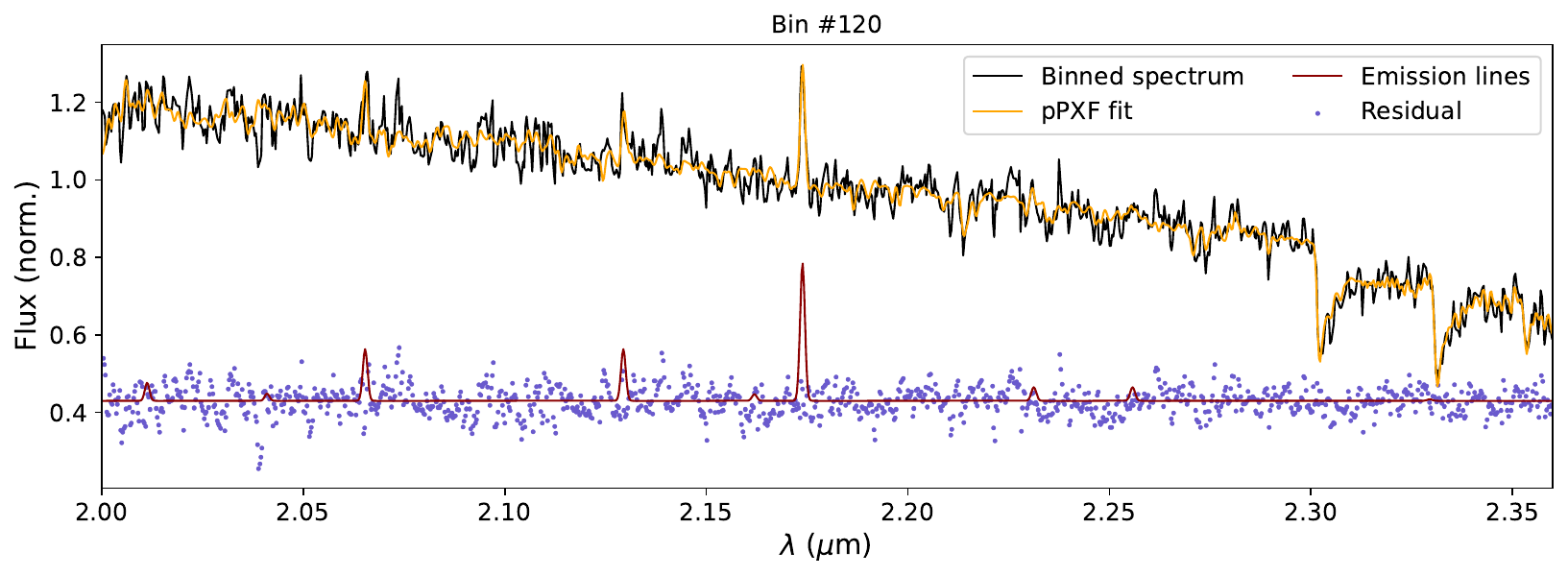}
    \caption{Example \textsc{pPXF} fit (orange) to a binned spectrum (black). The residual is shown as purple dots shifted to an arbitrary flux for visualisation. The red curve shows the best fitting emission lines.}
    \label{fig:co_bandheads}
\end{figure*}

\begin{figure*}
\centering
    \includegraphics[width=0.90\textwidth]{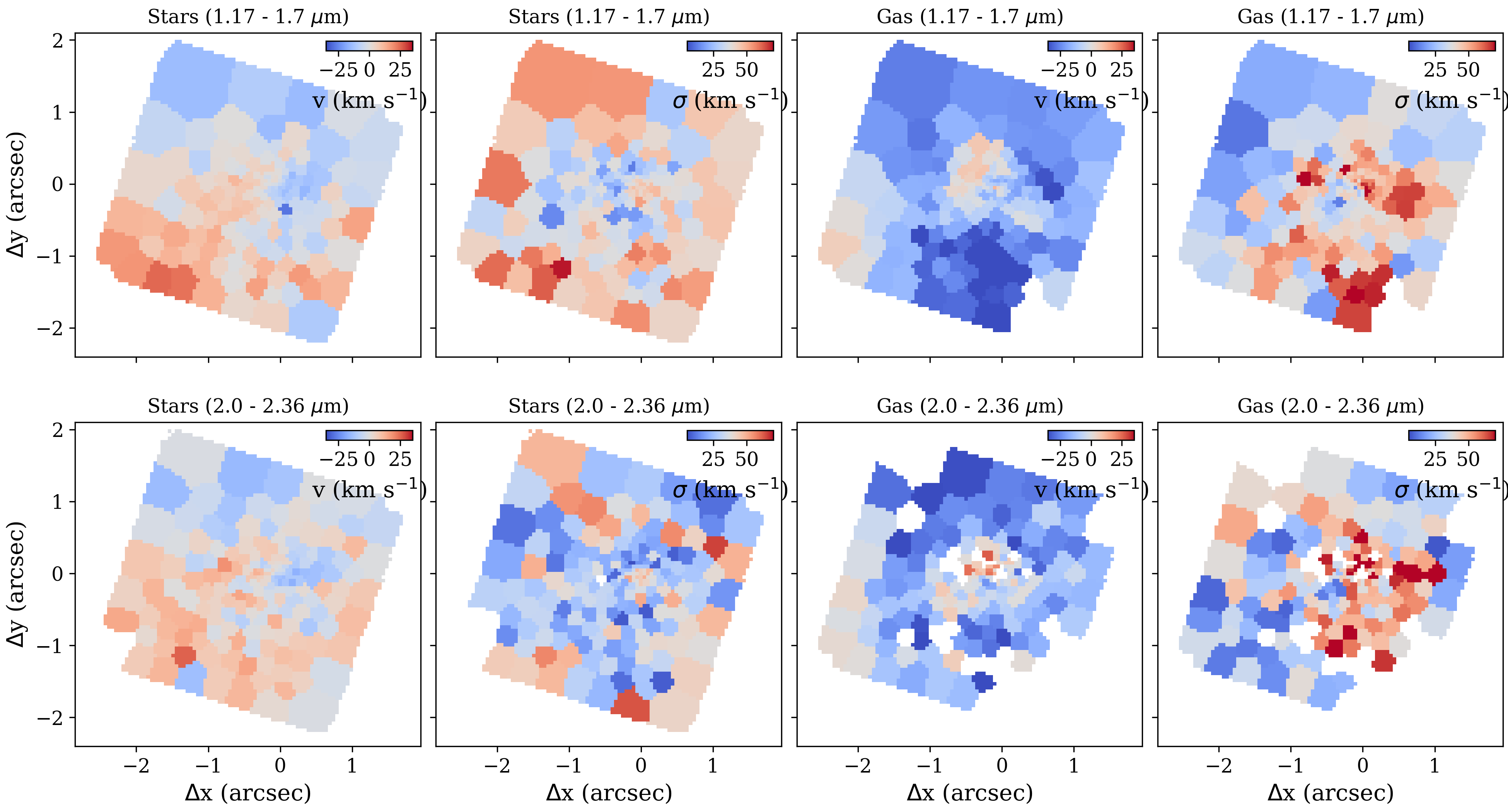}
    \caption{\textit{From left to right:} Stellar velocity and dispersion, gas velocity, and dispersion from the g140h grating (fitted between 1.17 - 1.7 $\mu$m, \textit{top}) and the g235h grating (fitted between 2.0 - 2.36 $\mu$m, \textit{bottom}).}
    \label{fig:co_bandhead_kinematics}
\end{figure*}

In Sect. \ref{sect:velocities}, we present kinematic maps based on full spectrum fitting in the wavelength range from 1.17 to 1.7 $\mu$m. As an alternative, we show here results obtained by fitting the CO bandheads at 2.3 $\mu$m, a strong absorption feature originating in the atmosphere of late-type stars, which is often used to derive stellar kinematics because of its sharp band edge and relative insensitivity to the properties of the stellar population. For this reason, we spatially binned the G235H data cube to S/N = 50 and fitted the spectra between 2.0 and 2.36 $\mu$m (redder wavelengths can affected by the gap between NIRSpec detectors) with \textsc{pPXF} and the XSL SSP models. Figure \ref{fig:co_bandheads} shows an example for one of the bins. As can be seen, not only absorption lines are fitted, but also emission lines. In this wavelength range this mainly includes $H_2$ at 2.12 $\mu$m and 2.03 $\mu$m and Brackett $\gamma$ at 2.17 $\mu$m. In this example bin, those lines are well fitted, however as the emission line maps in Fig. \ref{fig:emission_line_maps} or the cluster spectra in Fig. \ref{fig:cluster_spectra} show, these lines can be rather weak in some regions. Obtaining gas kinematics from this wavelength range is therefore associated with larger uncertainties than the range from 1.17 to 1.7 $\mu$m.
For this reason, the resulting gas kinematic maps (see Fig. \ref{fig:co_bandhead_kinematics}) from the 2.0 to 2.36 $\mu$m are not very reliable. Nevertheless, the stellar velocity map is very similar to the one derived in the g140h grating. The stellar velocity dispersion map also agrees in the central region, but tends to give low dispersions in the other bins. We believe that this is an effect of the lower spectral resolution in the 2.0 - 2.36 $\mu$m wavelength range ($R \sim 2500$, while the 1.17 to 1.7 $\mu$m range has a median $R \sim 2760$).

\section{Posterior distribution plots from SED modelling}
\label{app:corner_plots}
\begin{figure*}
    \centering
    \includegraphics[width=0.80\textwidth]{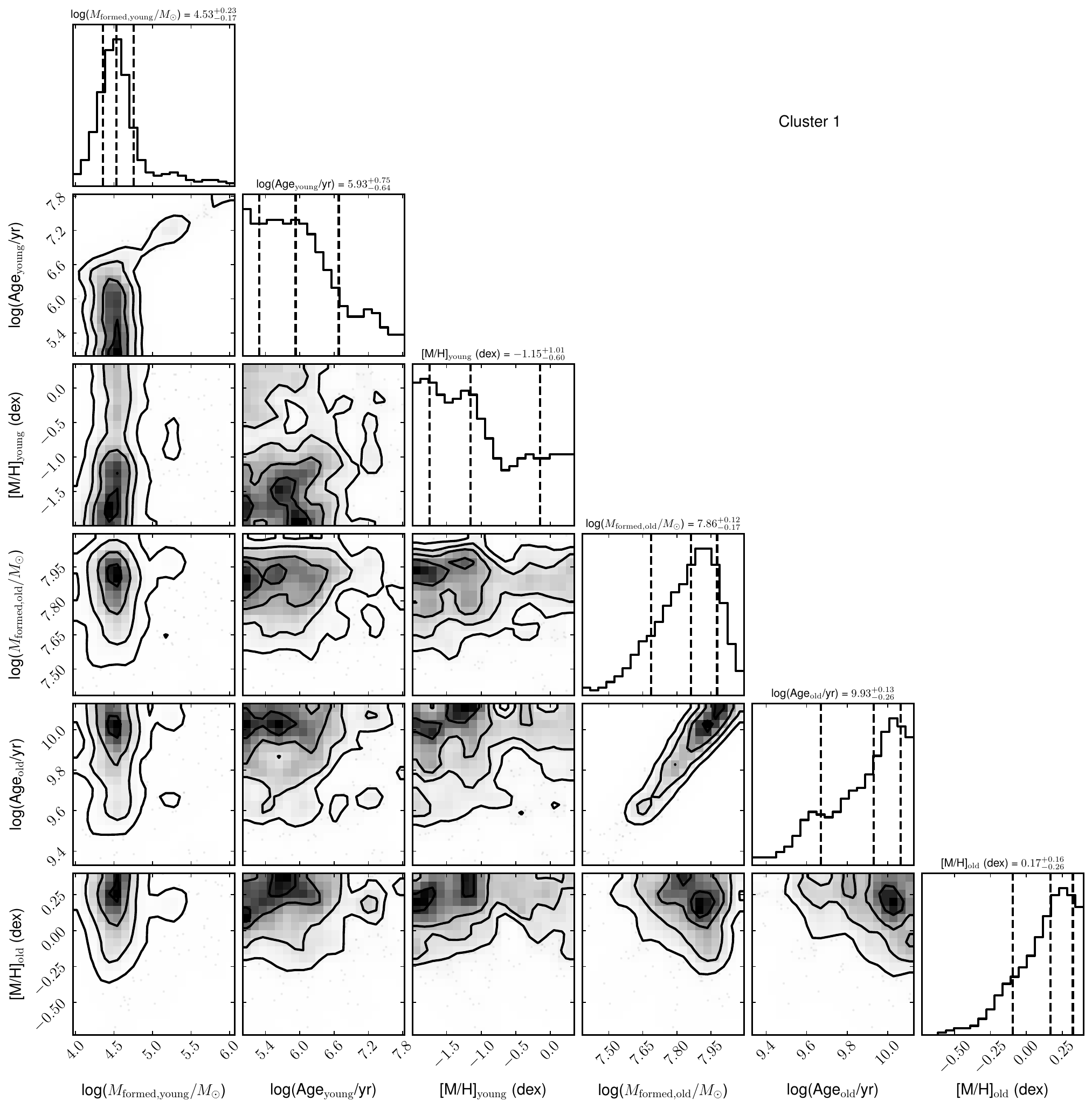}
    \caption{Corner plot showing the posterior distributions of the \textsc{bagpipes} fit to the fluxes of cluster 1. The dashed lines show the 16th, 50th, and 84th percentiles of the distributions. As described in Sect. \ref{sect:sed_fitting}, this cluster requires two populations to explain both the UV flux and the optical and NIR slopes. Only free parameters are shown here.}
    \label{fig:corner1}
\end{figure*}

\begin{figure*}
    \centering
    \includegraphics[width=0.80\textwidth]{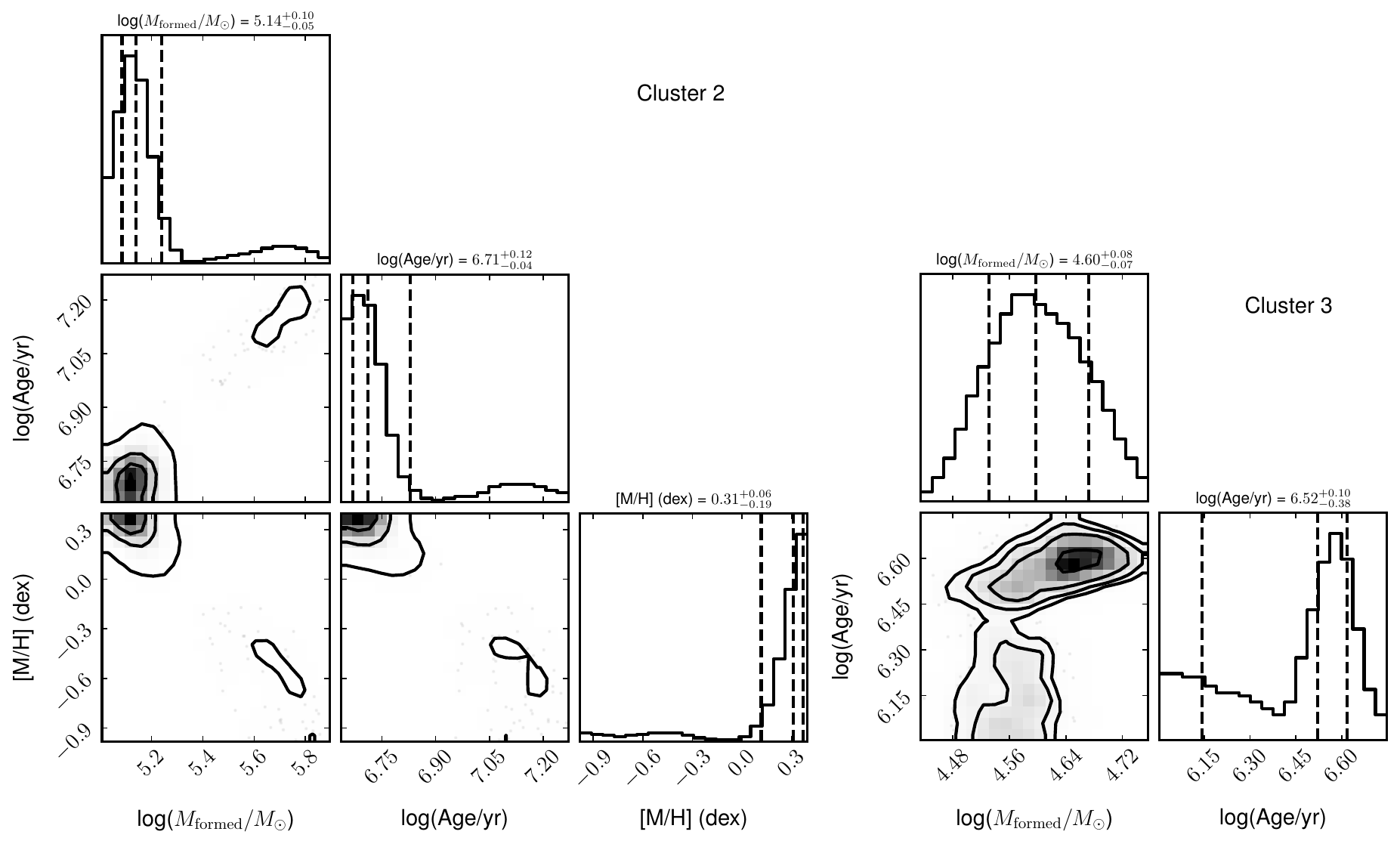}
    \caption{Corner plots showing the posterior distributions from the fits to cluster 2 (\textit{left}) and cluster 3 (\textit{right}). As cluster 3 is much fainter, we restricted its metallicity to the same of cluster 2.}
    \label{fig:corner2}
\end{figure*}

Figures \ref{fig:corner1} and \ref{fig:corner2} shows the posterior distributions from the \textsc{bagpipes} SED fitting of the three clusters, plotted with \textsc{corner} \citep{corner}. We only show the free parameters used in the fitting. The mass weighted ages and metallicities are reported in Table \ref{tab:cluster_properties}.
The mass formed in these figures ($M_\text{formed}$) refers to the stellar mass at formation time and differs from the current stellar mass due to stellar evolution. This is particularly true for old stellar populations such as old population in cluster 1.

\end{appendix}

\end{document}